\documentclass[11pt]{amsart}%
\usepackage{graphicx}
\usepackage{amscd}
\usepackage{amsbsy}
\usepackage{amsmath}
\usepackage{amstext}%
\usepackage{amsfonts}%
\usepackage{amssymb}

\theoremstyle{plain}

\newtheorem{definition}{Definition}

\numberwithin{equation}{section}

\makeindex

\begin{document}
\title[Quantum Crytography]{A Talk on Quantum Cryptography\\or\\How Alice Outwits Eve\\Version 1.6}
\author{Samuel J. Lomonaco, Jr.}
\address{University of Maryland Baltimore County\\
Baltimore, MD \ 21250}
\email{Lomonaco@UMBC.EDU}
\urladdr{http://www.umbc.edu/\symbol{126}lomonaco}
\thanks{Partially supported by ARL Contract \#DAAL01-95-P-1884, ARO Grant
\#P-38804-PH-QC, the Computer Security Division of NIST, and the L-O-O-P
Fund.~\ This paper is a revised version of a paper published in ``Coding
Theory, and Cryptography: From Geheimscheimschreiber and Enigma to Quantum
Theory,'' (edited by David Joyner), Lecture Notes in Computer Science and
Engineering, Springer-Verlag, 1999 (pp. 144-174). \ It has been reproduced
with the permission of Springer-Verlag.}
\date{January 27, 2001}
\keywords{}

\begin{abstract}
Alice and Bob wish to communicate without the archvillainess Eve eavesdropping
on their conversation. Alice, decides to take two college courses, one in
cryptography, the other in quantum mechanics. \ During the courses, she
discovers she can use what she has just learned to devise a cryptographic
communication system that automatically detects whether or not Eve is up to
her villainous eavesdropping. Some of the topics discussed are Heisenberg's
Uncertainty Principle, the Vernam cipher, the BB84 and B92 cryptographic
protocols. The talk ends with a discussion of some of Eve's possible
eavesdropping strategies, opaque eavesdropping, translucent eavesdropping, and
translucent eavesdropping with entanglement.\medskip

\end{abstract}
\maketitle
\tableofcontents

\section{Preface}

\bigskip

\subsection{The Unique Contribution of Quantum Cryptography}

Before beginning our story, I'd like to state precisely what is the unique
contribution of quantum cryptography.

\bigskip

\begin{itemize}
\item[ ] Quantum cryptography provides a new mechanism enabling the parties
communicating with one another to:
\[
\fbox{Automatically Detect Eavesdropping}%
\]
Consequently, it provides a means for determining when an encrypted
communication has been compromised.
\end{itemize}

\bigskip

\subsection{A Note to the Reader}

This paper is based on an invited talk given at the Conference on Coding
theory, Cryptology, and Number Theory held at the US Naval Academy in
Annapolis, Maryland in October of 1998. \ It was also given as an invited talk
at the Quantum Computational Science Workshop held in conjunction with the
Frontiers in Computing Conference in Annapolis, Maryland in February of 1999,
at a Bell Labs Colloquium in Murray Hill, New Jersey in April of 1999, at the
Security and Technology Division Colloquium of NIST in Gaithersburg, Maryland,
and at the Quantum Computation Seminar at the U.S. Naval Research Labs in
Washington, DC.

My objective in creating this paper was to write it exactly as I had given the
talk. \ But ... \ Shortly after starting this manuscript, I succumbed to the
temptation of greatly embellishing the story that had been woven into the
original talk. \ I leave it to the reader to decide whether or not this
detracts from or enhances the paper.

\section{Introduction}

We begin our crypto drama with the introduction of two of the main characters,
\textbf{Alice }$\heartsuit$and$\heartsuit$ \textbf{Bob}, representing
respectively the \textbf{sender} and the \textbf{receiver}. \ As in every
drama, there is a triangle. \ The triangle is completed with the introduction
of the third main character, the archvillainess \textbf{Eve}, representing the
\textbf{eavesdropper}. \ 

Our story begins with Alice and Bob attending two different universities which
are unfortunately separated by a great distance. \ Alice would like to
communicate with Bob without the ever vigilant Eve eavesdropping on their
conversation. \ In other words, how can Alice talk with Bob while at the same
time preventing the evil Eve from listening in on their conversation?

\section{A Course on Classical Cryptography}

\subsection{Alice's enthusiastic decision}

\bigskip

Hoping to find some way out of her dilemma, Alice elects to take a course on
cryptography, Crypto 351 taught by Professor Shannon with guest lecturers
Diffie, Rivest, Shamir, and Adleman. \ Alice thinks to herself, ``Certainly
this is a wise choice. \ It is a very applied course, and surely relevant to
the real world. \ Maybe I will learn enough to outwit Eve?''\bigskip

\subsection{Plaintext, ciphertext, key, and ... Catch 22}

\bigskip

Professor Shannon begins the course with a description of classical
cryptographic communication systems, as illustrated in Fig. 1. \ Alice, the
sender, encrypts her \textbf{plaintext} $P$ into \textbf{ciphertext} $C$ using
a \textbf{secret key} $K$ which she shares only with Bob, and sends the
ciphertext $C$ over an \textbf{insecure channel} on which the evil Eve is ever
vigilantly eavesdropping. \ Bob, the receiver, receives the ciphertext $C$,
and uses the secret key $K$, shared by him and Alice only, to decrypt the
ciphertext $C$ into plaintext $P$.%

\begin{center}
\includegraphics[
trim=0.000000in 0.655883in 0.000000in 0.000000in,
height=1.7971in,
width=4.2583in
]%
{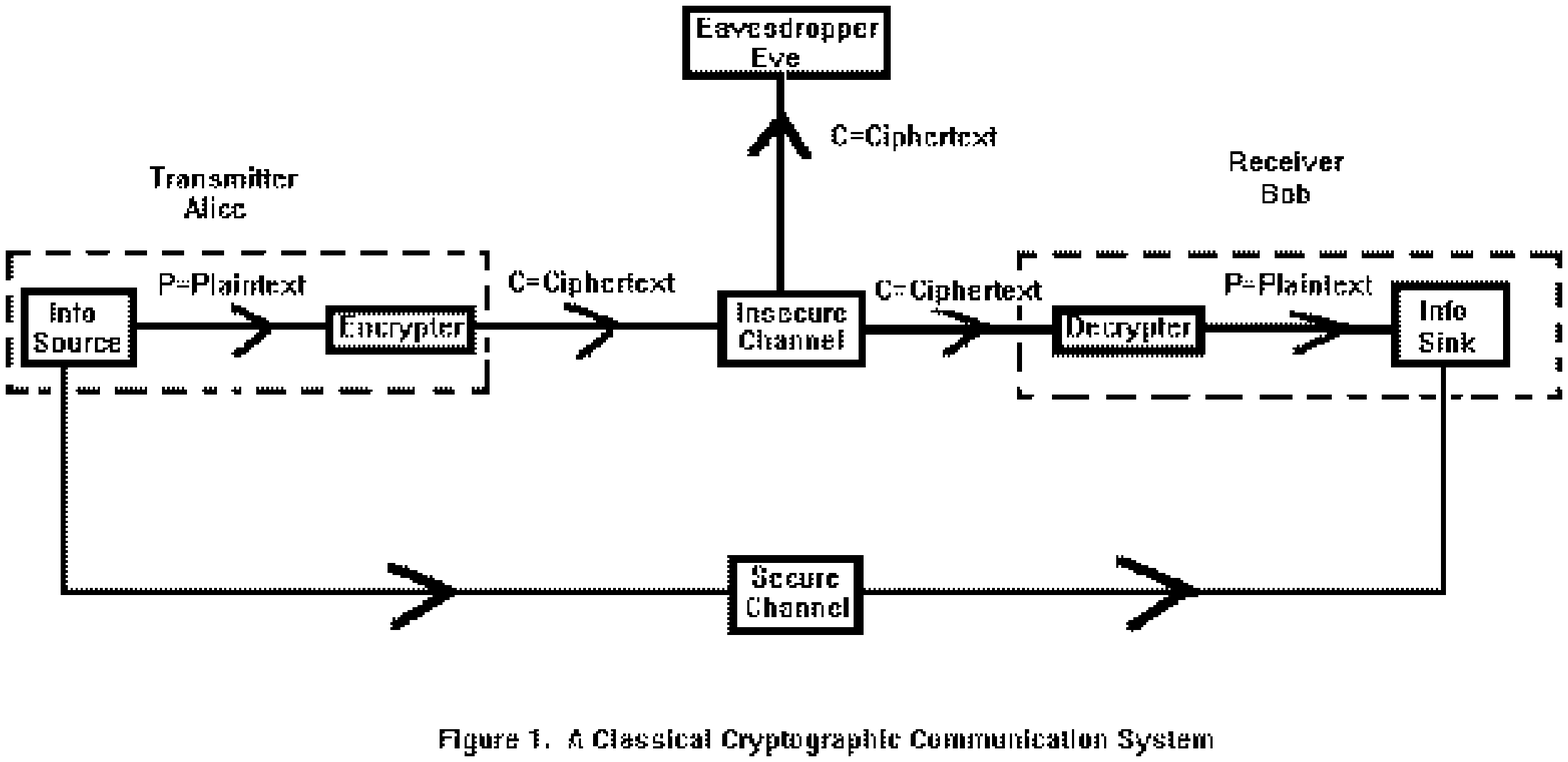}%
\\
Figure 1. A classical cryptographic communication system.
\end{center}

What is usually not mentioned in the description of a classical cryptographic
communication system is that Alice and Bob must first communicate over a
\textbf{secure channel} to establish a secret key $K$ shared only by Alice and
Bob before they can communicate in secret over the insecure channel. \ Such a
channel could consist, for example, of a trusted courier, wearing a trench
coat and dark sunglasses, transporting from Alice to Bob a locked briefcase
chained to his wrist. \ In other words, we have the famous Catch 22 of
classical cryptography, namely:

\bigskip

\begin{itemize}
\item[ ] \textbf{Catch 22.} \ There are perfectly good ways to communicate in
secret, provided we can communicate in secret ...
\end{itemize}

\bigskip

Professor Shannon then goes on to discuss the different types of classical
communication security.

\bigskip

\subsection{Practical Secrecy}

A cryptographic communication system is \textbf{practically secure} if the
encryption scheme can be broken after $X$ years, where $X$ is determined by
one's security needs and by existing technology. \ Practically secure
cryptographic systems have existed since antiquity. \ One example would be the
Caesar cipher used by Julius Caesar during the during the Gallic wars, a
cipher that was difficult for his opponents to break at that time, but easily
breakable by today's standards. \ A modern day example of a practically secure
classical cryptographic system is the digital encryption standard (DES) which
has just recently been broken\footnote{Tim O'Reilly \ and the Electronic
Frontier Foundation have constructed a computing device for \$250,000 which
does an exhaustive key search on DES in 4.5 days\cite{O'Reilly1}. \ See also
\cite{Biham1} and \cite{Matsui1}. \ As far as I know, triple DES has not been
broken.}. \ For this and many other reasons, DES is to be replaced by a more
practically secure classical encryption system, the Advanced Encryption
Standard (AES). \ In turn, AES will be replaced by an even more secure
cryptographic system should the advances in technology ever challenge its security.

\bigskip

\subsection{Perfect Secrecy}

A cryptographic communication is said to be \textbf{perfectly secure} if the
ciphertext $C$ gives no information whatsoever about the plaintext $P$, even
when the design of the cryptographic system is known. \ In mathematical terms,
this can be stated succinctly with the equation:
\[
PROB(P\ |\ C)=PROB(P)\text{.}%
\]
In other words, the probability of plaintext $P$ given ciphertext $C$, written
$PROB(P|C)$, is equal to the probability of the plaintext $P$.

\bigskip

An example of a perfectly secure classical cryptographic system is the
\textbf{Vernam Cipher}, better known as the \textbf{One-Time-Pad}. \ The
plaintext $P$ is a binary sequence of zeroes and ones, i.e.,
\[
P=P_{1},P_{2},P_{3},\ \ldots\ P_{n},\ \ldots\
\]
The secret key $K$ consists of a totally random binary sequence of the same
length, i.e.,
\[
K=K_{1},K_{2},K_{3},\ \ldots\ ,\ K_{n},\ \ldots\
\]
\ The ciphertext $C$ is the binary sequence
\[
C=C_{1},C_{2},C_{3},\ \ldots\ C_{n},\ \ldots
\]
obtained by adding the sequences $P$ and $K$ bitwise modulo $2$, i.e.,
\[
C_{i}=P_{i}+K_{i}\operatorname{mod}2\text{ \qquad for \qquad}i=1,2,3,\ \ldots
\]
For example,
\[%
\begin{array}
[c]{rrcrrrr}%
P\ \ \  & = & 0110\quad0101\quad1101 &  &  &  & \\
K\ \ \  & = & 1010\quad1110\quad0100 &  &  &  & \\
\_\_\_\_\_\_\_\_\_\_\_ &  & \_\_\_\_\_\_\_\_\_\_\_\_\_\_\_\_\_\_\_ &  &  &  &
\\
C=P\oplus K\ \ \  & = & 1100\quad1011\quad1001 &  &  &  &
\end{array}
\]

\bigskip

This cipher is perfectly secure if key $K$ is totally random and shared only
by Alice and Bob. \ It is easy to encode with the key $K$. \ If, however, one
succumbs to the temptation of using the same key $K$ to encode two different
plaintexts $P^{(1)}$ and $P^{(2)}$ into ciphertexts $C^{(1)}$ and $C^{(2)}$,
then the cipher system immediately changes from a perfectly secure cipher to
one that is easily broken by even the most amateur cryptanalyst. \ For,
$C^{(1)}\oplus C^{(2)}=P^{(1)}\oplus P^{(2)}$ is easily breakable because of
the redundancy that is usually present in plaintext.

\bigskip

The only problem with the one-time-pad is that long bit sequences must be sent
over a secure channel before it can be used. \ This once again leads us to the
Catch 22 of classical cryptography, i.e.,

\bigskip

\begin{itemize}
\item[ ] \textbf{Catch 22.} \ There are perfectly good ways to communicate in
secret, provided we can communicate in secret ...
\end{itemize}

\bigskip

... and to the:

\bigskip

\begin{itemize}
\item \textbf{Key Problem 1.} \ \emph{Catch 22}: A secure means of
communicating key is needed.\footnote{Hired trench coats are exorbitantly
expensive and time consuming.}\bigskip
\end{itemize}

Finally, there are two other key problems in classical cryptography in need of
a solution, namely:

\bigskip

\begin{itemize}
\item \textbf{Key Problem 2.} \ \emph{Authentication}: Alice needs to
determine with certainty that she is actually talking to Bob, and not to an
impostor such as Eve.

\item \textbf{Key Problem 3.} \ \emph{Intrusion Detection}: Alice needs a
means of determining whether or not Eve is eavesdropping.
\end{itemize}

\bigskip

In summary, we have the following checklist for classical cryptographic systems:%

\[
\fbox{$%
\begin{array}
[c]{lr}%
\text{\underline{\textbf{Check List}} for Classical Crypto Systems} & \\
& \\
\qquad\blacksquare\text{ Catch 22 Solved?} & \qquad\qquad\text{\textbf{NO}}\\
& \\
\qquad\blacksquare\text{ Authentication?} & \qquad\qquad\text{\textbf{NO}}\\
& \\
\qquad\blacksquare\text{ Intrusion Detection?} & \qquad\qquad\text{\textbf{NO}%
}%
\end{array}
$}%
\]

\bigskip

\subsection{Computational Security}

Relatively recently in the history of cryptography, Diffie and Hellman
\cite{Diffie1}, \cite{Diffie2} suggested a new type cryptographic secrecy. \ A
cipher is said to be \textbf{computationally secure} if the computational
resources required to break it exceed anything possible now and into the
future. \ For example, a cipher would be computationally secure if the number
of bits of computer memory required to break it were greater than the number
of atoms in the universe, or if the computational time required to break it
exceeded the age of the universe. Cryptographic systems can be created in such
a way that it is computationally infeasible to find the decryption key $D$
even when the encryption key $E$ is known. \ To create such a cryptographic
system, all one would need is a trap-door function $f$.

\bigskip

\begin{definition}
A function $f$ is a \textbf{trap-door function} if

\begin{itemize}
\item[1)] $f$ is easy to compute, i.e., polynomial time computable, and

\item[2)] Given the function $f$, the inverse function $f^{-1}$can not be
computed from $f$ in polynomial time, i.e., such a computation is
superpolynomial time, intractable, or worse.
\end{itemize}
\end{definition}%

\begin{center}
\fbox{\includegraphics[
height=1.9692in,
width=2.8003in
]%
{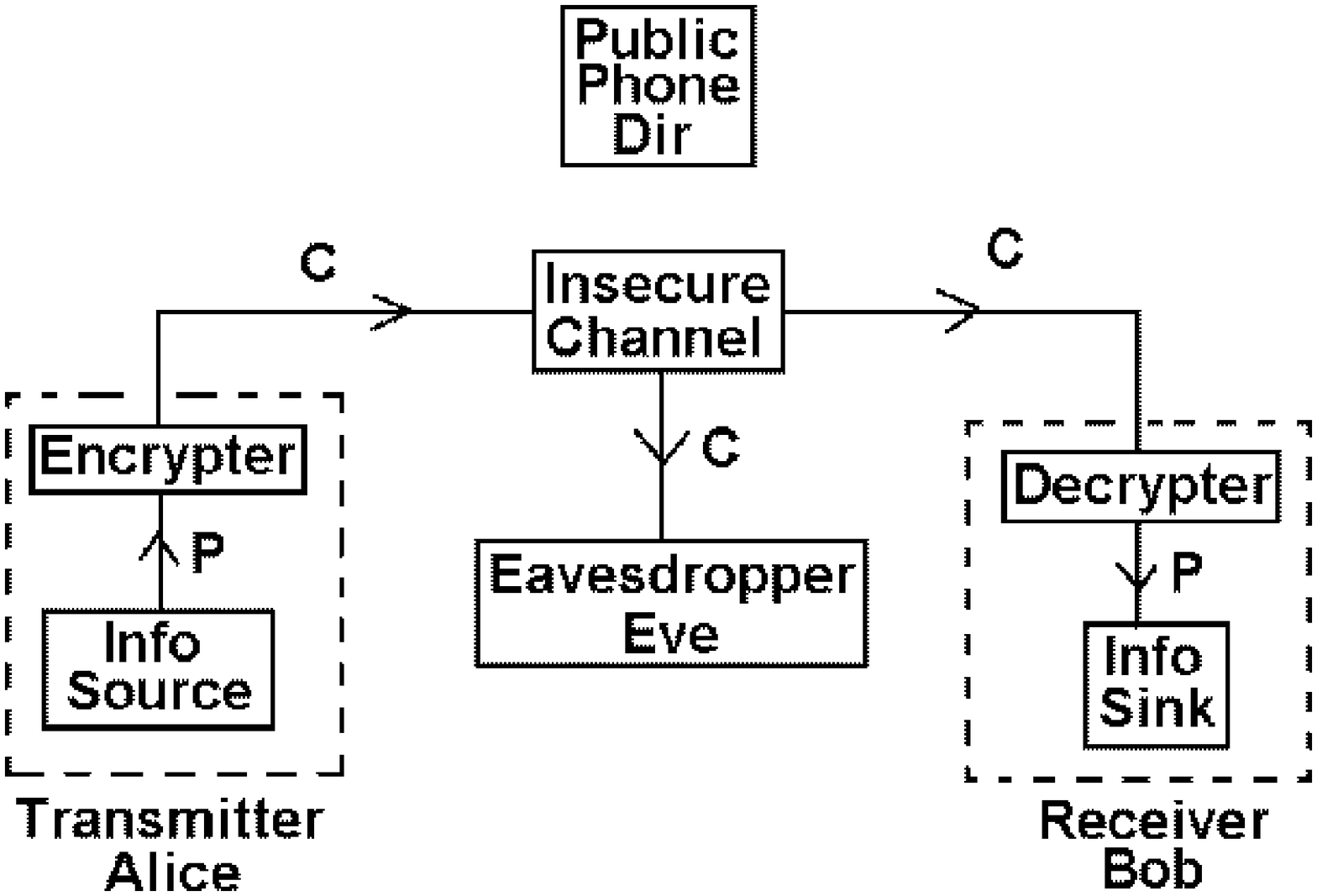}%
}\\
Figure 2. A public key cryptographic communication system
\end{center}

\bigskip

A trap-door function $E$ can be used to create a \textbf{public key
cryptographic system} as illustrated in Fig.2. \ All parties who wish to
communicate in secret should choose their own trap-door function $E$ and place
it in a \textbf{public directory}, the ``\textbf{yellow pages},'' for all the
world to see. But they should keep their decryption key $D=E^{-1}$ secret.
\ Since $E$ is a trap-door function, it is computationally infeasible for
anyone to use the publicly known $E$ to find the decryption key $D$. So $D$ is
secure in spite of the fact that its inverse $E$ is publicly known.

\bigskip

If Alice wishes to send a secret communication to Bob, she first looks up in
the yellow pages Bob's encryption key $E_{B}$, encrypts her plaintext $P$ with
Bob's encryption key $E_{B}$ to produce ciphertext $C=E_{B}(P)$, and then
sends the ciphertext $C$ over a public channel. \ Bob receives the ciphertext
$C$, and decrypts it back into plaintext $P=D_{B}(C)$ using his secret
decryption key $D_{B}$.

\bigskip

Alice can even do more than this. \ She can authenticate, i.e., sign her
encrypted communication to Bob so that Bob knows with certitude that the
message he received actually came from Alice and not from an Eve masquerading
as Alice. \ Alice can do this by encrypting her signature $\mathcal{ALICE}$
using her secret decryption key $D_{A}$ into $D_{A}(\mathcal{ALICE})$. She
then encrypts plaintext $P$ plus her signature $D_{A}(\mathcal{ALICE})$ using
Bob's publicly known encryption key $E_{B}$ to produce the signed ciphertext
$C_{\mathcal{S}}=E_{B}(P+D_{A}(\mathcal{ALICE}))$, and then sends her signed
ciphertext $C_{\mathcal{S}}$ over the public channel to Bob. \ Bob can then
decrypt the message as he did before to produce the signed plaintext
$P+D_{A}(\mathcal{ALICE})$. \ Bob can verify Alice's digital signature
$D_{A}(\mathcal{ALICE})$ by looking up Alice's encryption key $E_{A}$ in the
``yellow pages,'' and using it to find her signature $E_{A}(D_{A}%
(\mathcal{ALICE}))=\mathcal{ALICE}$. In this way, he authenticates that Alice
actually sent the message because only she knows her secret decryption key.
\ Hence, only she could have signed the plaintext.\footnote{Because of the
need for brevity, we have not discussed all the subtleties involved with
digital signatures. \ For example, for more security, Alice should add a time
stamp and some random symbols to her signature. \ For more information on
digital signatures, please refer to one of the standard references such as
\cite{Menezes1}.}

\bigskip

The RSA\ cryptographic system is believed to be one example of a public key
cryptographic system. \ There are many public software implementations of RSA,
e.g., PGS (Pretty Good Security).

\bigskip

Thus, besides solving the authentication problem for cryptography, public key
cryptographic systems appear also to solve the Catch 22 of cryptography.
\ However, frequently the encryption and decryption keys of a public key
cryptographic system are managed by a central key bank. \ In this case, the
Catch 22 problem is still there. \ For that reason, we have entered `MAYBE' in
the summary given below.

\bigskip%

\[
\fbox{$%
\begin{array}
[c]{ll}%
\text{\underline{\textbf{Check List}} for PKS} & \\
& \\
\qquad\blacksquare\text{ Catch 22 Solved?} & \qquad\qquad\text{\textbf{MAYBE}%
}\\
& \\
\qquad\blacksquare\text{ Authentication?} & \qquad\qquad\text{\textbf{YES}}\\
& \\
\qquad\blacksquare\text{ Intrusion Detection?} & \qquad\qquad\text{\textbf{NO}%
}%
\end{array}
$}%
\]

\section{A Course on Quantum Mechanics}

\subsection{Alice's Reluctant Decision}

In spite of Alice's many intense efforts to avoid taking a course in quantum
mechanics, she was finally forced by her university's General Education
Requirements (GERs) to register for the course Quantum 317, taught by
Professor Dirac with guest lecturers Feynman, Bennett, and Brassard. \ She did
so reluctantly. \ ``After all,'' she thought, ``Certainly this is an insane
requirement. \ Quantum mechanics is not applied. \ It's too theoretical to be
relevant to the real world. \ Ugh! \ But I do want to graduate.''

\bigskip

\subsection{The Classical World -- Introducing the Shannon Bit}

Professor Dirac began the course with a brief introduction to the classical
world of information. In particular, Alice was introduced to the classical
Shannon Bit, and shown that he/she/it is a very decisive individual. \ The
Shannon Bit is either $0$ or $1$, but by no means both at the same time. \ 

``Hmm ... ,'' she thought, ``I bet that almost everyone I know is gainfully
employed because of the Shannon Bit.'' \ 

The professor ended his brief discussion of the Shannon Bit by mentioning that
there is one of its properties that we take for granted. I.e., it can be
copied. \ 

\bigskip

\subsection{The Quantum World -- Introducing the Qubit}

Next Professor Dirac switched to the mysterious world of the quantum. \ He
began by introducing the runt of the Bit clan, i.e., the Quantum Bit,
nicknamed \textbf{Qubit}. \ He began by showing the class a small dot, i.e., a
quantum dot. \ In fact it was so small that Alice couldn't see it at all. \ He
promptly pulled out a microscope\footnote{This is a most unusual microscope!},
and projected a large image on a screen for the entire class to view. \ 

\bigskip

Professor Dirac went on to say, ``In contrast to the decisive classical
Shannon Bit, the Qubit is a very indecisive individual. \ It is both $0$ and
$1$ at the same time! \ Moreover, unlike the Shannon Bit, the Qubit cannot be
copied because of the no cloning theorem of Dieks, Wootters, and
Zurek\cite{Dieks1}\cite{Wootters1}. \ Qubits are very slippery characters,
exceedingly difficult to deal with.''

\bigskip

``One example of a qubit is a spin $\frac{1}{2}$ particle which can be in a
spin-up state $\left|  1\right\rangle $ which we label as $1$, in a spin-down
state $\left|  0\right\rangle $ which we label as $0$, or in a
\textbf{superposition} of these states, which we interpret as being both $0$
and $1$ at the same time.'' \ (The term ``superposition'' will be explained
shortly.) \ 

\bigskip

``Another example of a qubit is the polarization state of a photon. \ A photon
can be in a vertically polarized state $\left|  \updownarrow\right\rangle $.
We assign a label of $1$ to this state. \ It can be in a horizontally
polarized state $\left|  \leftrightarrow\right\rangle $. \ We assign a label
of $0$ to this state. Or, it can be in a superposition of these states. In
this case, we interpret its state as representing both $0$ and $1$ at the same
time.'' \ 

\bigskip

``Anyone who has worn polarized sunglasses should be familiar with the
polarization states of the photon. \ Polarized sunglasses eliminate glare
because they let through only vertically polarized light while filtering out
the horizontally polarized light that is reflected from the road.''

\bigskip

\subsection{Where do qubits live?}

\bigskip

But where do qubits live? \ They live in a Hilbert space $\mathcal{H}$. By a
Hilbert space, we mean:

\begin{definition}
A \textbf{Hilbert Space} is a vector space over the complex numbers
$\mathbb{C}$ together with an inner product
\[
\left\langle \quad,\quad\right\rangle :\mathcal{H}\times\mathcal{H}%
\longrightarrow\mathbb{C}%
\]
such that

\begin{itemize}
\item[1)] $\left\langle u_{1}+u_{2},v\right\rangle =\left\langle
u_{1},v\right\rangle +\left\langle u_{2},v\right\rangle $ for all $u_{1}%
,u_{2},v\in\mathcal{H}$

\item[2)] $\left\langle u,\lambda v\right\rangle =\left\langle \lambda
u,v\right\rangle $ for all $u,v\in\mathcal{H}$ and $\lambda\in\mathbb{C}$

\item[3)] $\left\langle u,v\right\rangle ^{\ast}=\left\langle v,u\right\rangle
$ for all $u,v\in\mathcal{H}$ , where the superscript `$\ast$' denotes complex conjugation.

\item[4)] For every Cauchy sequence $u_{1},u_{2},u_{3},\ \ldots\ $\ in
$\mathcal{H}$,
\[
\lim_{n\rightarrow\infty}u_{n}\text{ exists and lies in }\mathcal{H}%
\]
\end{itemize}
\end{definition}

\bigskip

In other words, a Hilbert space is a vector space over the complex numbers
$\mathbb{C}$ with a sequilinear inner product in which sequences that should
converge actually do converge to points in the space.\bigskip

\subsection{Some Dirac notation -- Introducing kets}

\bigskip

The elements of $\mathcal{H}$ are called \textbf{kets}, and will be denoted
by
\[
\left|  label\right\rangle \text{ ,}%
\]
where `$|$' and `$>$' are left and right delimiters, and `$label$' denotes any
label, i.e., name, we wish to assign to a ket.

\bigskip

\subsection{Finally, a definition of a qubit}

\bigskip

So finally, we can define what is meant by a qubit.

\begin{definition}
A \textbf{qubit} is a ket (state) in a two dimensional Hilbert space
$\mathcal{H}$.
\end{definition}

\bigskip

Thus, if we let $\left|  0\right\rangle $ and $\left|  1\right\rangle $ denote
an arbitrary orthonormal basis of a two dimensional Hilbert space
$\mathcal{H}$, then each qubit in $\mathcal{H}$ can be written in the form
\[
\left|  qubit\right\rangle =\alpha_{0}\left|  0\right\rangle +\alpha
_{1}\left|  0\right\rangle
\]
where $\alpha_{0}$, $\alpha_{1}\in\mathbb{C}$. \ Since any scalar multiple of
a ket represents the same state of an isolated quantum system, we can assume,
without loss of generality, that $\left|  qubit\right\rangle $ is a ket of
unit length, i.e., that
\[
\left|  \alpha_{0}\right|  ^{2}+\left|  \alpha_{1}\right|  ^{2}=1
\]

The above qubit is said to be in a \textbf{superposition} of the states
$\left|  0\right\rangle $ and $\left|  1\right\rangle $. \ This is what we
mean when we say that a qubit can be simultaneously both $0$ and $1$. However,
if the qubit is observed it immediately ``makes a decision.'' \ It ``decides''
to be $0$ with probability $\left|  \alpha_{0}\right|  ^{2}$ and $1$ with
probability $\left|  \alpha_{1}\right|  ^{2}$. Some physicists call this the
\textbf{``collapse''} of the wave function\footnote{It is very difficult, if
not impossible, to find two physicists who agree on the subject of quantum
measurement. \ The phrase ``collapse of the wave function'' immediately
engenders a ``war cry'' in most physicists. \ For that reason, ``collapse'' is
enclosed in quotes.}.

\bigskip

\subsection{More Dirac notation -- Introducing bras and bra-c-kets}

Given a Hilbert space $\mathcal{H}$, let
\[
\mathcal{H}^{\ast}=Hom(\mathcal{H},\mathbb{C})
\]
denote the set of all linear maps from $\mathcal{H}$ to $\mathbb{C}$. \ Then
$\mathcal{H}^{\ast}$ is actually a Hilbert space, called the \textbf{dual}
Hilbert space of $\mathcal{H}$, with scalar product and vector sum defined
by:
\[
\left\{
\begin{array}
[c]{lcll}%
\left(  \lambda\cdot f\right)  \left(  \left|  \Psi\right\rangle \right)  &
= & \lambda\left(  f(\left|  \Psi\right\rangle )\right)  \text{, } & \text{for
all }\lambda\in\mathbb{C}\text{ and for all }f\in\mathcal{H}^{\ast}\\
&  &  & \\
\left(  f_{1}+f_{2}\right)  \left(  \left|  \Psi\right\rangle \right)  & = &
f_{1}\left(  \left|  \Psi\right\rangle \right)  +f_{2}\left(  \left|
\Psi\right\rangle \right)  \text{, } & \text{for all }f_{1},f_{2}%
\in\mathcal{H}^{\ast}%
\end{array}
\right.
\]

We call the elements of $\mathcal{H}^{\ast}$ \textbf{bra}'s, and denote them
as:
\[
\left\langle label\right|
\]

We can now define a bilinear map
\[
\mathcal{H}^{\ast}\times\mathcal{H}\longrightarrow\mathbb{C}%
\]
by
\[
\left(  \left\langle \Psi_{1}\right|  \right)  \left(  \left|  \Psi
_{2}\right\rangle \right)  \in\mathbb{C}%
\]
since bra $\left\langle \Psi_{1}\right|  $ is a complex valued function of
kets. We denote this product more simply as
\[
\left\langle \Psi_{1}\right.  |\left.  \Psi_{2}\right\rangle
\]
and call it the \textbf{Bra-c-Ket} (or \textbf{bracket}) of bra $\left\langle
\Psi_{1}\right|  $ and ket $\left|  \Psi_{2}\right\rangle $.

\bigskip

Finally, the bracket induces a dual correspondence\footnote{This is true for
finite dimensional Hilbert spaces. \ It is more subtle for infinite
dimensional Hilbert spaces.} between $\mathcal{H}$ and $\mathcal{H}^{\ast}$,
i.e.,
\[
\left|  \Psi_{2}\right\rangle \overset{D.C.}{\longleftrightarrow}\left\langle
\Psi_{1}\right|
\]

\bigskip

\subsection{Activities in the quantum world -- Unitary transformations}

\bigskip

All ``activities'' in the quantum world are linear transformations
\[
U:\mathcal{H}\longrightarrow\mathcal{H}%
\]
from the Hilbert space $\mathcal{H}$ into itself, called \textbf{unitary
transformations} (or, \textbf{unitary operators}). If we think of linear
transformations as matrices, then a \textbf{unitary transformation} $U$ is a
square matrix of complex numbers such that
\[
\overline{U}^{T}U=I=U\overline{U}^{T}%
\]
where $\overline{U}^{T}$ denotes the matrix obtained from $U$ by conjugating
all its entries and then transposing the matrix. \ We denote $\overline{U}%
^{T}$ by $U^{\dagger}$, and refer to it as the \textbf{adjoint} of $U$. \ 

\bigskip

Thus, an ``activity'' in the quantum world would be, for example, a unitary
transformation $U$ that carries a state ket $\left|  \Psi_{0}\right\rangle $
at time $t=0$ to a state ket $\left|  \Psi_{1}\right\rangle $ at time $t=1$,
i.e.,
\[
U:\left|  \Psi_{0}\right\rangle \longmapsto\left|  \Psi_{1}\right\rangle
\]

\subsection{Observables in quantum mechanics -- Hermitian operators}

In quantum mechanics, what does an observer observe? \ 

\bigskip

All \textbf{observables} in the quantum world are linear transformations
\[
\mathcal{O}:\mathcal{H}\longrightarrow\mathcal{H}%
\]
from the Hilbert space $\mathcal{H}$ into itself, called \textbf{Hermitian
operators} (or, \textbf{self-adjoint operators}). If we think of linear
transformations as matrices, then a \textbf{Hermitian operator} $\mathcal{O}$
is a square matrix of complex numbers such that
\[
\overline{\mathcal{O}}^{T}=\mathcal{O}%
\]
where $\overline{\mathcal{O}}^{T}$ again denotes the matrix obtained from
$\mathcal{O}$ by conjugating all its entries, and then transposing the matrix.
\ As before, we denote $\overline{\mathcal{O}}^{T}$ by $\mathcal{O}^{\dagger}%
$, and refer to it as the \textbf{adjoint} of $\mathcal{O}$. \ 

\bigskip

Let $\left|  \varphi_{i}\right\rangle $ denote the eigenvectors, called
\textbf{eigenkets}, of an observable $\mathcal{O}$, and let $a_{i}$ denote the
corresponding eigenvalue, i.e.,
\[
\mathcal{O}:\left|  \varphi_{i}\right\rangle =a_{i}\left|  \varphi
_{i}\right\rangle
\]
In the cases we consider in this talk, the eigenkets form an orthonormal basis
of the underlying Hilbert space $\mathcal{H}$.

\bigskip

Finally, we can answer our original question, i.e.,
\[
\fbox{What does an observer observe?}%
\]

\bigskip

Let us suppose that we have a physical device $M$ that is so constructed that
it measures an observable $\mathcal{O}$, and that we wish to use $M$ to
measure a quantum system which just happens to be in a quantum state $\left|
\Psi\right\rangle $. We assume $\left|  \Psi\right\rangle $ is a ket of unit
length. The quantum state $\left|  \Psi\right\rangle $ can be written as a
linear combination of the eigenkets of $\mathcal{O}$, i.e.,
\[
\left|  \Psi\right\rangle =%
{\displaystyle\sum}
\alpha_{i}\left|  \varphi_{i}\right\rangle
\]
When we use the device $M$ to measure $\left|  \Psi\right\rangle $, we observe
the eigenvalue $a_{i}$ with probability $p_{i}=\left|  \alpha_{i}\right|
^{2}$, and in addition, after the measurement the quantum system has
``collapsed'' into the state $\left|  \varphi_{i}\right\rangle $. \ Thus, the
outcome of a measurement is usually random, and usually has a lasting impact
on the state of the quantum system.

\bigskip

We can use Dirac notation to write down an expression for the average observed
value. \ Namely, the \textbf{averaged observed value} is given by the
expression $\left\langle \Psi\right|  \left(  \mathcal{O}\left|
\Psi\right\rangle \right)  $, which is written more succinctly as
$\left\langle \Psi\right|  \mathcal{O}\left|  \Psi\right\rangle $, or simply
as $\left\langle \mathcal{O}\right\rangle $.

\subsection{The Heisenberg uncertainty principle -- A limitation on what we
can actually observe}

\bigskip

There is, surprisingly enough, a limitation of what can be observed in quantum mechanics.

\bigskip

Two observables $A$ and $B$ are said to be \textbf{compatible} if they
commute, i.e., if
\[
AB=BA\text{.}%
\]
Otherwise, they are said to be \textbf{incompatible}.

\bigskip

Let $\left[  A,B\right]  $, called the \textbf{commutator} of $A$ and $B$,
denote the expression
\[
\left[  A,B\right]  =AB-BA
\]
In this notation, two operators $A$ and $B$ are compatible if and only if
$\left[  A,B\right]  =0$. \ Finally, let
\[
\triangle A=A-\left\langle A\right\rangle
\]

The following principle is one expression of how quantum mechanics places
limits on what can be observed:

\bigskip

\noindent\textbf{Heisenberg's Uncertainty Principle}\footnote{We have assumed
units have been chosen such that $\hslash=1$.}
\[
\left\langle \left(  \triangle A\right)  ^{2}\right\rangle \left\langle
\left(  \triangle B\right)  ^{2}\right\rangle \geq\frac{1}{4}\left\|
\left\langle \left[  A,B\right]  \right\rangle \right\|  ^{2}%
\]
where $\left\langle \left(  \triangle A\right)  ^{2}\right\rangle
=\left\langle \Psi\right|  \left(  \triangle A\right)  ^{2}\left|
\Psi\right\rangle $ is the \textbf{mean squared standard deviation} of the
observed eigenvalue, written in Dirac notation. It is a measure of the
uncertainty in $A$.

\bigskip

This if $A$ and $B$ are incompatible, i.e., do not commute, then, by measuring
$A$ more precisely, we are forced to measure $B$ less precisely, and vice
versa. \ We can not simultaneously measure both $A$ and $B$ to to unlimited
precision. Measurement of $A$ somehow has an impact on the measurement of $B$.

\subsection{Young's two slit experiment -- An example of Heisenberg's
uncertainty principle}

\bigskip

For the purpose of illustrating Heisenberg's Uncertainty Principle, Professor
Dirac wheeled out into the classroom a device to demonstrate Young's two slit
experiment. \ The device consisted of an electron gun which spewed out
electrons\footnote{The original Young's two slit experiment used photons
rather than electrons.} in the direction of a wall with two slits. The
electrons that managed to pass through the two slits then impacted on a
backstop which consisted of a $1600\times1200$ rectangular lattice of
extremely small counters. \ From the back of the backstop, all of the
1,920,000 tiny counters on the backstop were individually connected to a PC by
a cable consisting of a dense bundle of filaments. \ 

\bigskip

Professor Dirac pointed to the PC, and explained that the PC was setup to
display on the CRT's $1600\times1200$ pixel screen the individual running
total counts of all the backstop counters . \ He went on to say that the
intensity $P\left(  i,j\right)  $ of pixel $\left(  i,j\right)  $ on the
screen was proportional to the total number of electrons counted so far by the
counter at position $\left(  i,j\right)  $ in the backstop.

\bigskip

Professor Dirac proceeded to demonstrate what the device could actually do.
\ He turned on the electron gun, and turned down its intensity so low that the
probability of more than two electrons being emitted at the same time was
negligibly small.

\bigskip

His first experiment with the device was to cover slit 2, allowing the
incoming electrons to pass only through slit 1. \ He reset all the counters on
the backstop to zero, and then stepped back to let the students in his class
view the screen.

\bigskip

Initially nothing could be seen on the screen but blackness. \ However,
gradually an intensity pattern began to form on the screen. \ At first the
displayed pattern was indiscernible. \ But eventually it began to look like
the intensity pattern of a classical two dimensional Gaussian distribution.
\ He then pressed a key on his computer to show a three dimensional plot of
the intensity $P\left(  i,j\right)  $ as a surface in 3-space. \ Then with the
click of a mouse, he displayed a plot of the intensity $P\left(  i,j\right)  $
along the vertical line $j=800$ going down the center of the screen. \ The
plot was that of the bell shaped classical one dimensional Gaussian
distribution curve $P_{1}$, as shown in Fig. 3a. \ This was a clear indication
that the random impacts on the backstop were obeying the classical Gaussian distribution.

\bigskip

When he repeated the experiment with the slit 1 instead of slit 2 covered,
exactly the same pattern of a classical two dimensional Gaussian distribution
pattern was seen, but only this time shifted vertically down a short distance
on the screen. \ A plot of the intensity $P\left(  i,j\right)  $ of along the
vertical line $j=800$ going down the center of the screen is indicated by
curve $P_{2}$ shown in Fig. 3a.%

\begin{center}
\fbox{\includegraphics[
height=1.8637in,
width=2.9257in
]%
{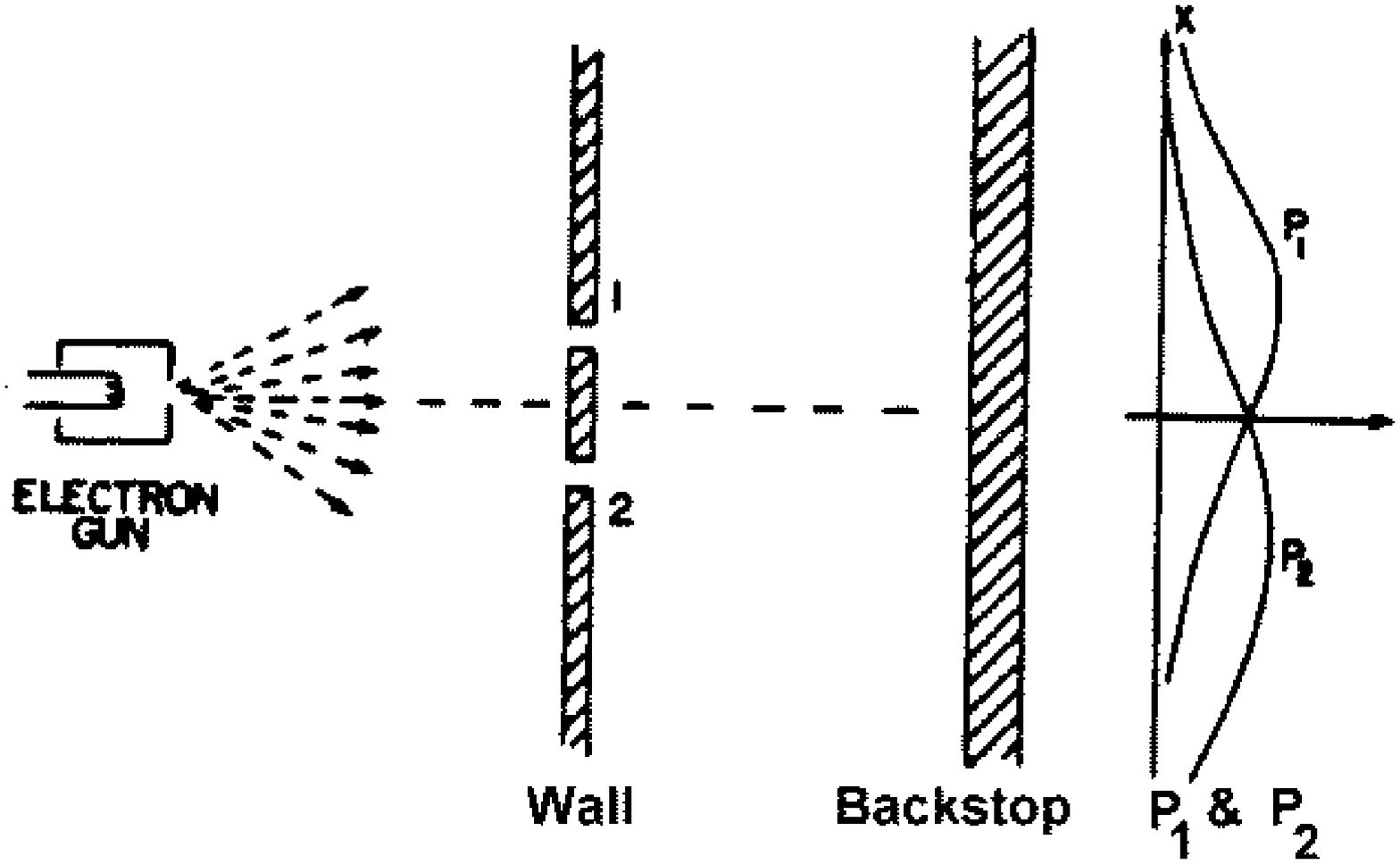}%
}\\
Figure 3a. Young's two slit experiment with one slit closed.
\end{center}

\bigskip

Professor Dirac then asked the students in the class what pattern they thought
would appear if he left both of the slits uncovered. \ Most of the class
responded by saying that the resulting light pattern would simply be the sum
of the two patterns, i.e., the bell shaped curve $P_{1}+P_{2}$, as illustrated
in Fig. 3c by the curve labeled $P_{12}^{\prime}$. \ Most of the class was
convinced that the two classical probability distributions would simply add,
as many of them had learned in the probability course Prob 323.

The remainder of the class stated quite emphatically that they did not care
what happened. \ What was being illustrated was far from an applied area, and
hence not relevant to their real world. \ Or so they thought ...

Professor Dirac smiled, and then proceeded to uncover both slits. \ What
appeared on the screen to almost everyone's surprise was not the pattern with
the bell shape $P_{1}+P_{2}$. \ It was instead a light pattern with a wavy
bell shaped curve, as illustrated by the curve $P_{12}$ in Fig. 3b.%

\begin{center}
\fbox{\includegraphics[
height=1.6051in,
width=2.8314in
]%
{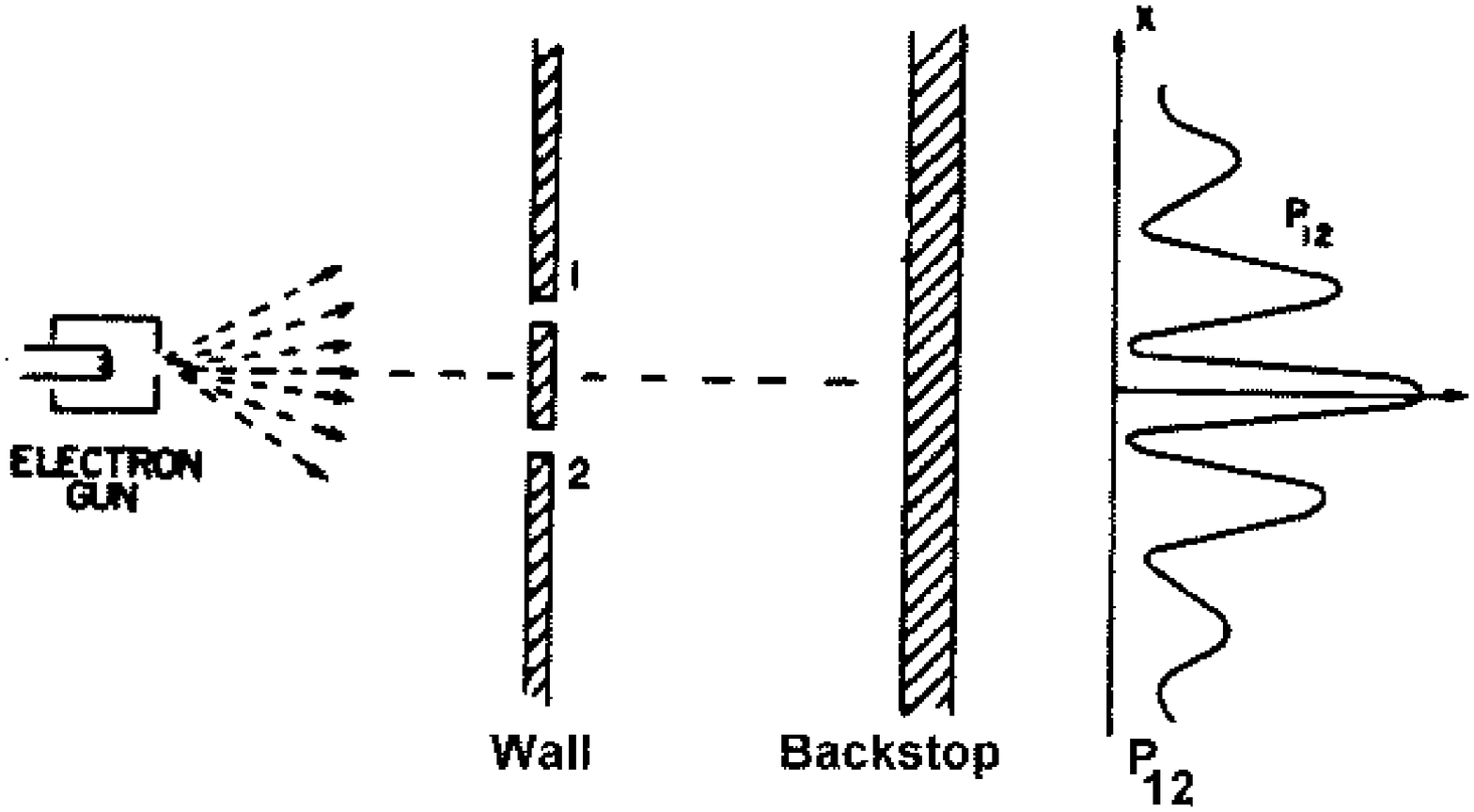}%
}\\
Figure 3b. Young's two slit experiment with both slits open. \
\end{center}

Professor Dirac explained, ``Something non-classical had occurred. \ Unlike
classical probabilities, the quantum probabilities (or more correctly stated,
the quantum amplitudes) had interfered with one another to produce an
interference pattern. \ In the dark areas, one finds destructive interference.
\ In the bright areas, one finds constructive interference.'' \ 

\bigskip

``Indeed, something non-classical is happening here.'' \ 

\bigskip

``Strangely enough, quantum mechanics is telling us that each electron is
actually passing through both slits simultaneously! \ It is as if each
electron were a wave and not a particle.''

\bigskip

\ 

``But what happens when we actually try to observe through which slit each
electron passes?''

\bigskip

Professor Dirac pulled out his trusty microscope\footnote{This is a most
unusual microscope.} to observe which of the two slits each electron passed
through. \ He reset all the backstop counters to zero, turned on the device,
and began observing through which slit each electron passed through. \ The
class was much surprised to find that the wavey interference pattern did not
appear on the screen this time. \ Instead, what appeared was the classical
intensity pattern all had initially expected to see in the first place, i.e.,
the intensity pattern of the bell shaped curve $P_{12}=P_{1}+P_{2}$, as shown
in Fig. 3c. \ 

\bigskip

''So we see that, when observed, the electrons act as particles and not as waves!''%

\begin{center}
\fbox{\includegraphics[
height=1.6449in,
width=3.1989in
]%
{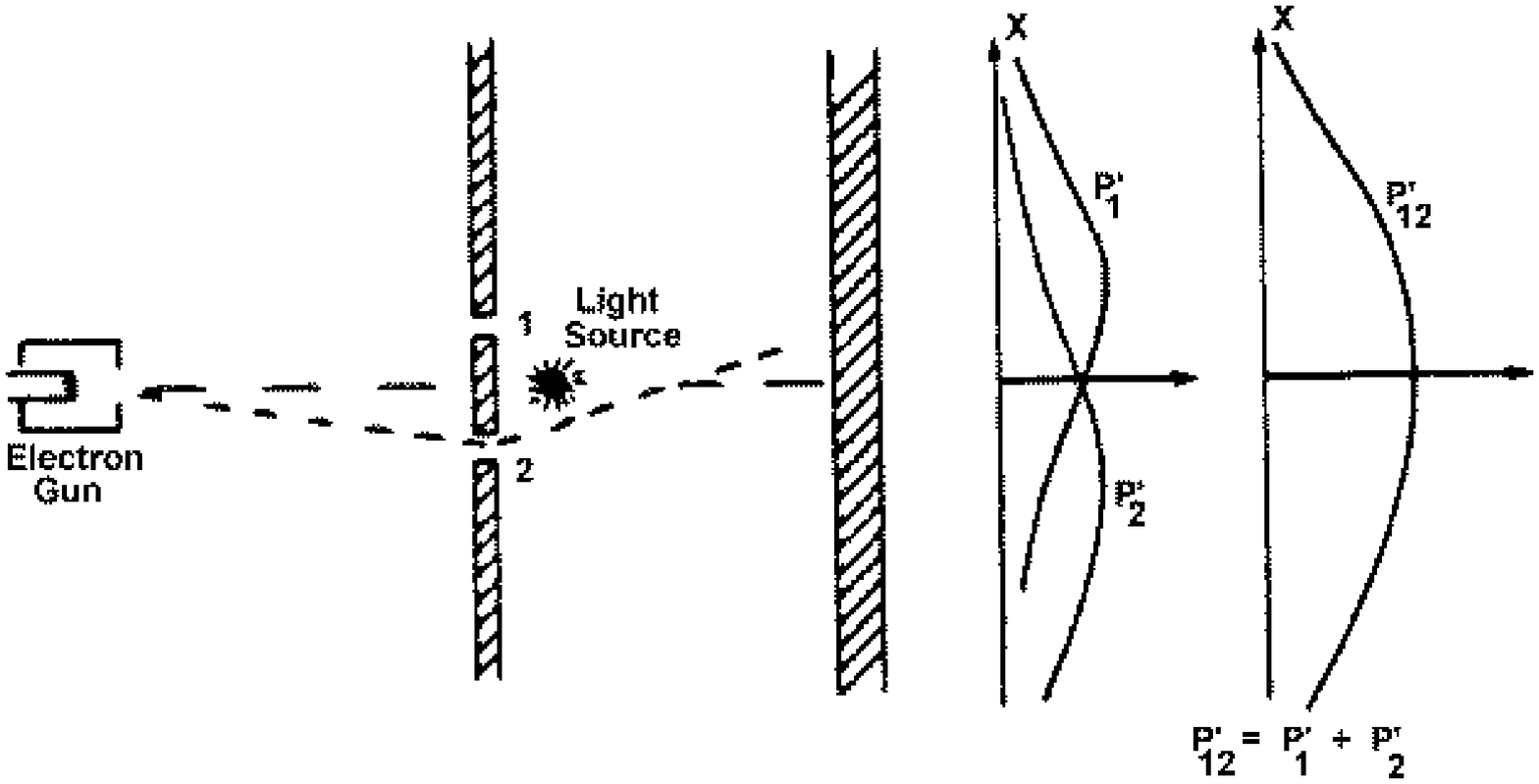}%
}\\
Figure 3c. \ Young's 2 slit experiment when the slit through which the
electron passes is determined by observation.
\end{center}

After a brief pause, Professor Dirac said, ``This is actually an example of
the Heisenberg Uncertainty Principle. \ We can see this as follows:''

``In the experiment, we are effectively observing two incompatible
observables, the position operator $X$ (i.e., which slit each electron passes
through) and the momentum operator $P$ (i.e., the momentum with which each
electron leaves the slitted wall.) \ When we observe the momentum $P$, the
interference pattern is present. \ But when we observe the position $X$, the
interference pattern vanishes. \ We can not observe position without
disturbing momentum, and vice versa.''

\section{The Beginnings of Quantum Cryptography}

\subsection{Alice has an idea}

After class on her way back to her dorm room, Alice began once again to
ruminate over her dilemma in regard to Bob and Eve. \ 

``If only her message to Bob were like the interference pattern in Young's two
slit experiment. \ Then, if the prying Eve were to observe which of the two
slits each of the electrons emerged from (i.e., `listen in'), Bob would know
of her presence. \ For, if Eve were observing the individual electrons as they
left the slits, the pattern on the screen would be distorted from the
beautiful wavy interference pattern in a direction toward the dull ugly
Gaussian distribution pattern. \ Bob would see this distortion, and thereby be
able to surmise that Eve was eavesdropping.''%

\begin{center}
\fbox{\includegraphics[
height=1.6916in,
width=2.6602in
]%
{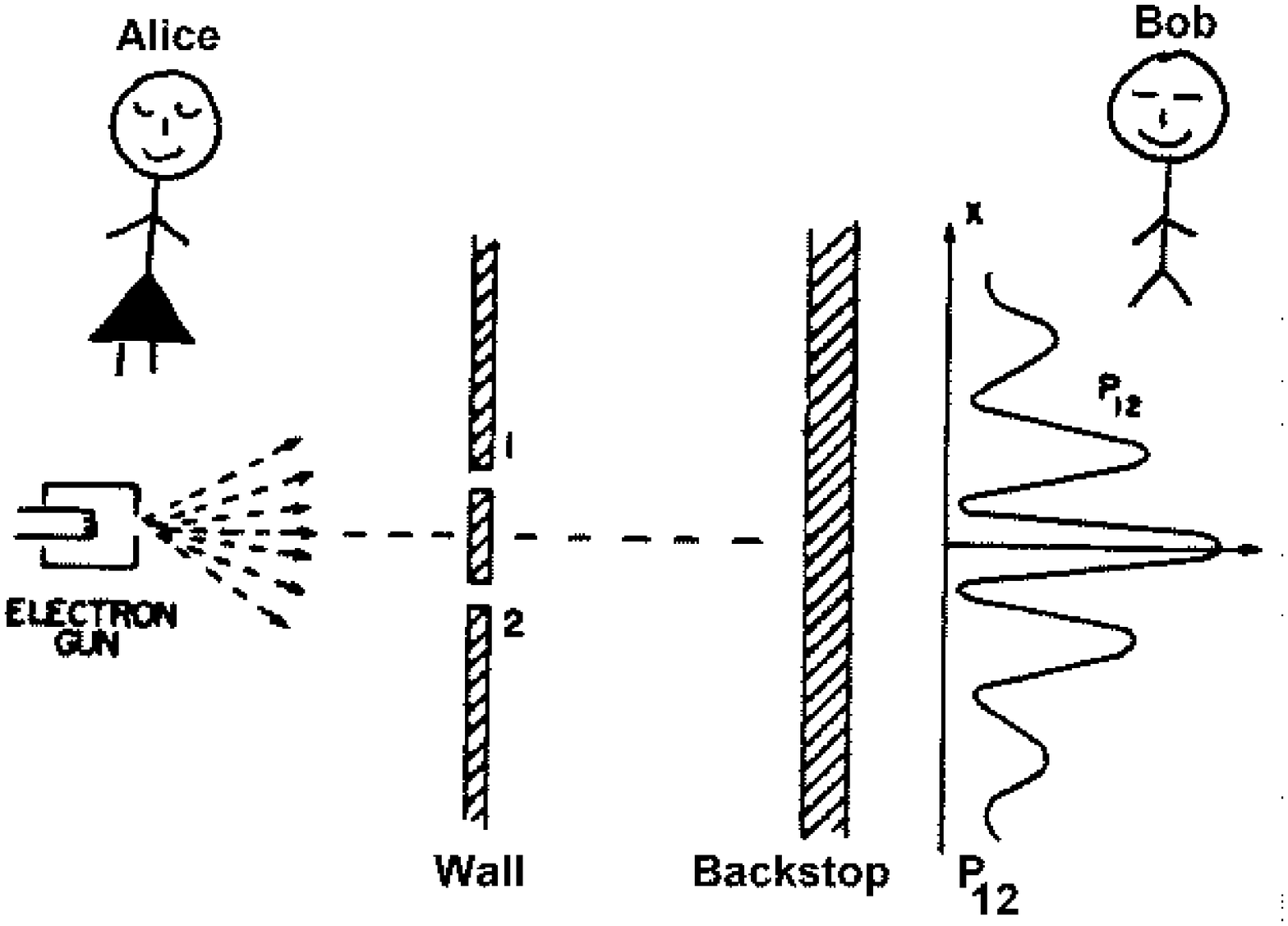}%
}\\
Figure 4a. Bob sees an interference pattern when Eve is not eavesdropping.
\end{center}
\begin{center}
\fbox{\includegraphics[
height=1.7045in,
width=2.6956in
]%
{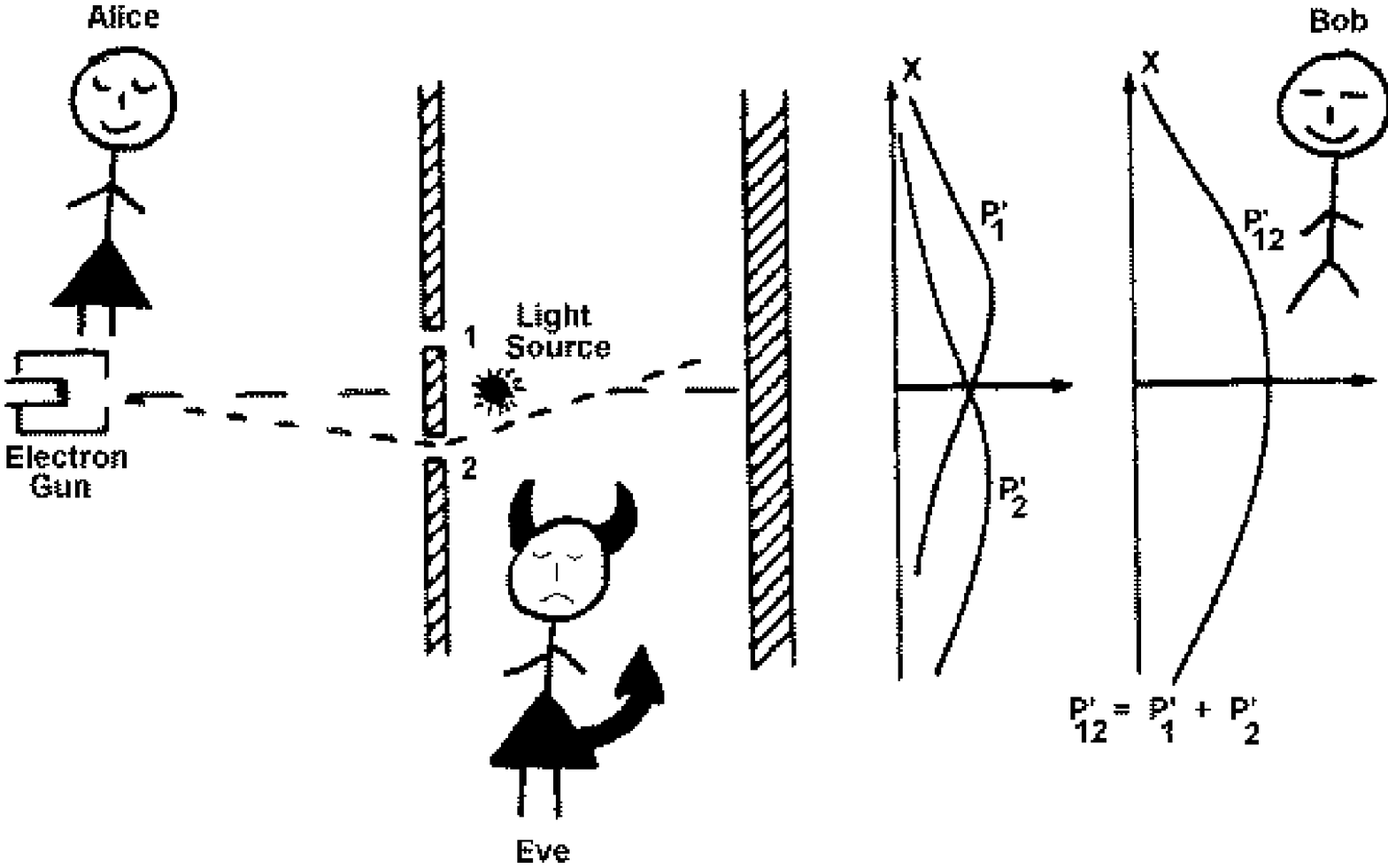}%
}\\
Figure 4b. Bob sees no interference pattern when Eve is eavesdropping.
\end{center}
\medskip

``This idea has possibilities. \ Maybe quantum mechanics is relevant after all!''

\bigskip

Her mind began to race. \ ``Perhaps something like Young's two slit experiment
could be used to communicate random key $K$? \ Then Bob could tell which key
had been compromised by an intruder such as Eve. \ But most importantly, he
could also surmise which key had not been compromised. \ Bob could then
communicate to me over the phone (or even over any public channel available
also to Eve) whether or not the key had been compromised, without, of course,
revealing the key itself. \ Any uncompromised key could then be employed to
send Bob a message by using the one-time-pad that was mentioned yesterday in
Crypto 351.''

``The beauty of this approach is that the one-time-pad is perfectly secure.
\ There is no way whatsoever that Eve could get any information about our
conversation. \ This would be true even if I used the campus radio station to
send my encrypted message.''\ 

``The evil Eve is foiled! \ Eureka! \ Contrary to student conventional wisdom,
both cryptography and quantum mechanics are relevant to the real world!''

\bigskip

``I have discovered a new kind of secrecy, i.e., quantum secrecy, which has
built-in detection of eavesdropping based on the principles of quantum
mechanics. \ I can hardly wait to tell Professor Dirac. \ She ran immediately
to his office.''

\bigskip

After listening to Alice's excited impromptu, and at times disjointed,
explanation, Professor Dirac suggested that she present her newly found
discoveries in his next class. \ Alice happily agreed to do so.

\subsection{Quantum secrecy -- The BB84 protocol without noise}

\bigskip

Two days later, after two sleepless but productive nights of work, Alice was
prepared for her presentation. \ She walked in the classroom for Quantum 317
carrying an overhead projector and a sizable bundle of transparencies.

\bigskip

After Professor Dirac had turned the large lecture hall over to her, she began
as follows:

\bigskip

``Let us suppose that I (Alice) would like to transmit a secret key $K$ to
Bob. \ Let us also suppose that someone by the name of Eve intends to make
every effort to eavesdrop on the transmission and learn the secret key.''

\bigskip

Wouldn't you know it. \ Eve just so happens to be sitting in the classroom!

\bigskip

``My objective today is to show you how the principles of quantum mechanics
can be used to build a cryptographic communication system in such a way that
the system detects if Eve is eavesdropping, and which also gives a guarantee
of no intrusion if Eve is not eavesdropping.''\bigskip

``A diagrammatic outline of the system I'm about to describe is shown on the
screen. (Please refer to Fig. 5.) \ Please note that the system consists of
two communication channels. \ One is a non-classical one-way quantum
communication channel, which I will soon describe.\ The other is an ordinary
run-of the-mill classical two-way public channel, such as a two-way radio
communication system. \ I emphasize that this classical two-way channel is
public, and open to whomever would like to listen in. \ For the time being, I
will assume that the two-way public channel is noise free.''\ %

\begin{center}
\fbox{\includegraphics[
height=1.7314in,
width=2.5425in
]%
{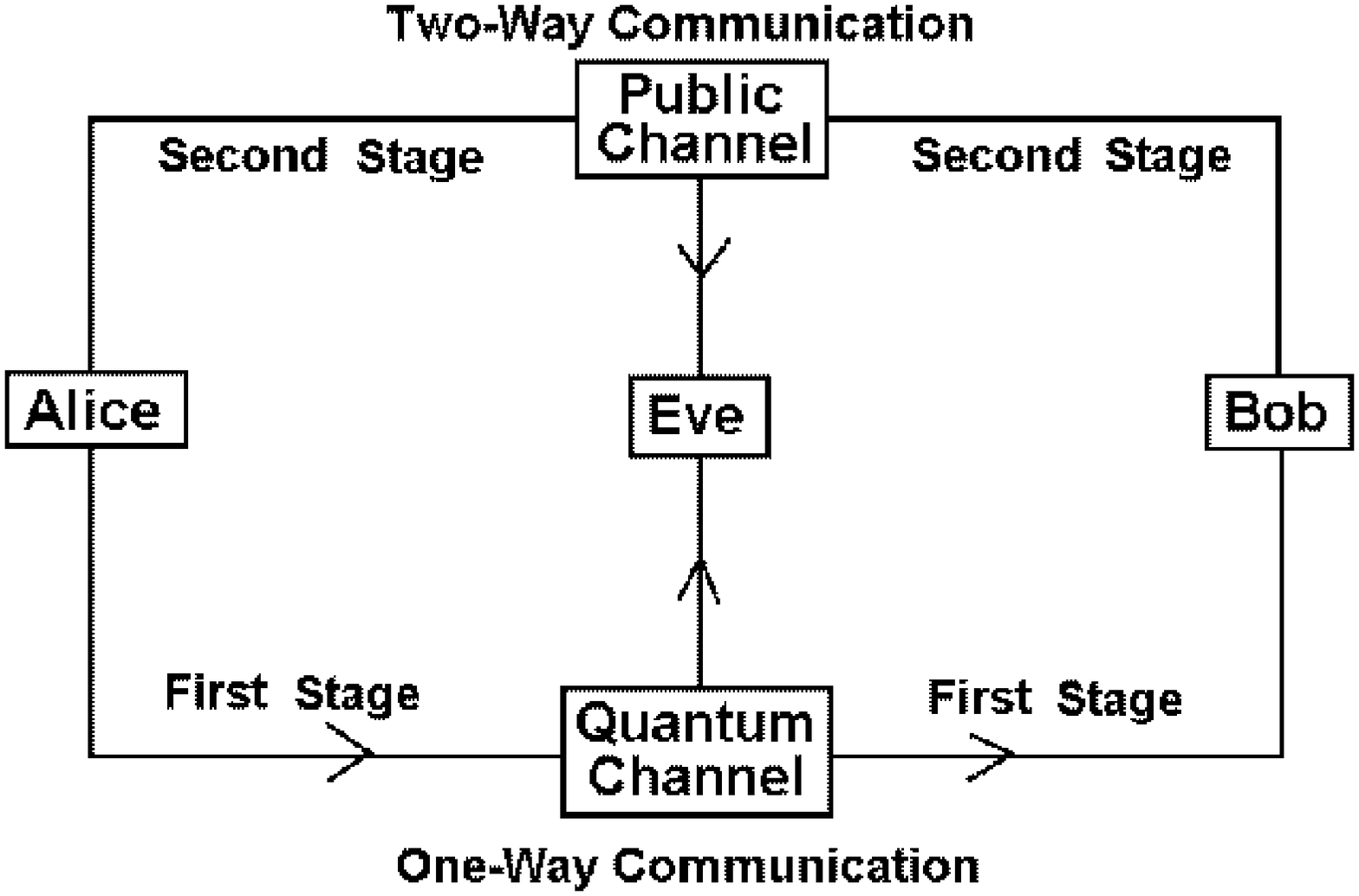}%
}\\
Figure 5. A quantum cryptographic communication system.
\end{center}

\bigskip

``I will now describe how the polarization states of the photon can be used to
construct a quantum one-way communication channel\footnote{Any two dimensional
quantum system such as a spin $\frac{1}{2}$ particle could be used.}.''

\bigskip

``From Professor Dirac's last lecture, we know that the polarization states of
a photon lie in a two dimensional Hilbert space $\mathcal{H}$. \ For this
space, there are many orthonormal bases. \ We will use only two for our
quantum channel.''

\bigskip

``The first is the basis consisting of the vertical and horizontal
polarization states, i.e., the kets $\left|  \updownarrow\right\rangle $ and
$\left|  \leftrightarrow\right\rangle $, respectively. \ We will refer to this
orthonormal basis as the \textbf{vertical/horizontal (V/H) basis}, and denote
this basis with the symbol `$\boxplus$.' ''

\bigskip

``The second orthonormal basis consists of the polarization states $\left|
\nearrow\right\rangle $ and $\left|  \nwarrow\right\rangle $, which correspond
to polarizations directions formed respectively by 45\% clockwise and
counter-clockwise rotations off from the vertical. \ We call this the
\textbf{oblique basis}, and denote this basis with the symbol `$\boxtimes$.' \ ''

\bigskip

``If I (Alice) decide to use the VH basis $\boxplus$ on the quantum channel,
then I will use the following \textbf{quantum alphabet}:
\[
\left\{
\begin{array}
[c]{ccc}%
``1" & = & \left|  \updownarrow\right\rangle \\
&  & \\
``0" & = & \left|  \leftrightarrow\right\rangle
\end{array}
\right.
\]
In other words, if I use this quantum alphabet on the quantum channel, I will
transmit a ``1'' to Bob simply by sending a photon in the polarization state
$\left|  \updownarrow\right\rangle $., and I will transmit a ``0'' by sending
a photon in the polarization state $\left|  \leftrightarrow\right\rangle $.''

\bigskip

``On the other hand, if I\ (Alice) decide to use the oblique basis $\boxtimes
$, then I will use the following \textbf{quantum alphabet}:
\[
\left\{
\begin{array}
[c]{ccc}%
``1" & = & \left|  \nearrow\right\rangle \\
&  & \\
``0" & = & \left|  \nwarrow\right\rangle
\end{array}
\right.  ,
\]
sending a ``1'' as a photon in the polarization state $\left|  \nearrow
\right\rangle $, and sending a ``0'' as a photon in the polarization state
$\left|  \nwarrow\right\rangle $.''

\bigskip

``I have chosen these two bases because the Heisenberg Uncertainty Principle
implies that observations with respect to the $\boxplus$ basis are
incompatible with observations with respect to the $\boxtimes$ basis. \ We
will soon see how this incompatibility can be translated into intrusion detection.''%

\begin{align*}
&
\begin{tabular}
[c]{lllllllllll}\hline
\multicolumn{1}{|l}{Alice} & \multicolumn{1}{|l}{$\boxplus$} &
\multicolumn{1}{|l}{$\boxtimes$} & \multicolumn{1}{|l}{$\boxtimes$} &
\multicolumn{1}{|l}{$\boxtimes$} & \multicolumn{1}{|l}{$\boxplus$} &
\multicolumn{1}{|l}{$\boxtimes$} & \multicolumn{1}{|l}{$\boxplus$} &
\multicolumn{1}{|l}{$\boxtimes$} & \multicolumn{1}{|l}{$\boxplus$} &
\multicolumn{1}{|l|}{$\boxtimes$}\\\hline
\multicolumn{1}{|l}{} & \multicolumn{1}{|l}{$\updownarrow$} &
\multicolumn{1}{|l}{$\nwarrow$} & \multicolumn{1}{|l}{$\nwarrow$} &
\multicolumn{1}{|l}{$\nearrow$} & \multicolumn{1}{|l}{$\updownarrow$} &
\multicolumn{1}{|l}{$\nwarrow$} & \multicolumn{1}{|l}{$\leftrightarrow$} &
\multicolumn{1}{|l}{$\nearrow$} & \multicolumn{1}{|l}{$\leftrightarrow$} &
\multicolumn{1}{|l|}{$\nearrow$}\\\hline
\multicolumn{1}{|l}{} & \multicolumn{1}{|l}{1} & \multicolumn{1}{|l}{0} &
\multicolumn{1}{|l}{0} & \multicolumn{1}{|l}{1} & \multicolumn{1}{|l}{1} &
\multicolumn{1}{|l}{0} & \multicolumn{1}{|l}{0} & \multicolumn{1}{|l}{1} &
\multicolumn{1}{|l}{0} & \multicolumn{1}{|l|}{1}\\\hline
& $\mid$ &  & $\mid$ &  &  &  &  & $\mid$ &  & $\mid$\\
& $\ast$ &  & $\ast$ &  &  &  &  & $\ast$ &  & $\ast$\\
& $\mid$ &  & $\mid$ &  &  &  &  & $\mid$ &  & $\mid$\\\hline
\multicolumn{1}{|l}{Bob} & \multicolumn{1}{|l}{$\boxtimes$} &
\multicolumn{1}{|l}{$\boxtimes$} & \multicolumn{1}{|l}{$\boxplus$} &
\multicolumn{1}{|l}{$\boxtimes$} & \multicolumn{1}{|l}{$\boxplus$} &
\multicolumn{1}{|l}{$\boxtimes$} & \multicolumn{1}{|l}{$\boxplus$} &
\multicolumn{1}{|l}{$\boxplus$} & \multicolumn{1}{|l}{$\boxplus$} &
\multicolumn{1}{|l|}{$\boxplus$}\\\hline
\multicolumn{1}{|l}{} & \multicolumn{1}{|l}{1} & \multicolumn{1}{|l}{0} &
\multicolumn{1}{|l}{1} & \multicolumn{1}{|l}{1} & \multicolumn{1}{|l}{1} &
\multicolumn{1}{|l}{0} & \multicolumn{1}{|l}{0} & \multicolumn{1}{|l}{0} &
\multicolumn{1}{|l}{0} & \multicolumn{1}{|l|}{0}\\\hline
&  &  &  &  & $\Updownarrow$ &  &  &  &  & \\\cline{2-11}%
Raw Key$\Rightarrow$ & \multicolumn{1}{|l}{} & \multicolumn{1}{|l}{0} &
\multicolumn{1}{|l}{} & \multicolumn{1}{|l}{1} & \multicolumn{1}{|l}{1} &
\multicolumn{1}{|l}{0} & \multicolumn{1}{|l}{0} & \multicolumn{1}{|l}{} &
\multicolumn{1}{|l}{0} & \multicolumn{1}{|l|}{}\\\cline{2-11}%
\end{tabular}
\\
&  \text{{\LARGE Fig. 6a. \ The BB84 protocol without\ Eve present (No
noise)}}%
\end{align*}

\bigskip

Alice and Bob now communicate with one another using a two stage protocol,
called the \textbf{BB84 protocol}\cite{Bennett1}. (Please refer to Figs.
6a\ and 6b.) \ 

\bigskip

In stage 1, Alice creates a random sequence of bits, which she sends to Bob
over the quantum channel using the following protocol:

\bigskip

\noindent\textbf{Stage 1 protocol: Communication over a quantum channel}\bigskip

\begin{itemize}
\item[Step 1.] Alice flips a fair coin to generate a random sequence
$S_{Alice}$ of zeroes and ones. \ This sequence will be used to construct a
secret key shared only by Alice and Bob.\bigskip

\item[Step 2.] For each bit of the random sequence, Alice flips a fair coin
again to choose at random one of the two quantum alphabets. \ She then
transmits the bit as a polarized photon according to the chosen alphabet.\bigskip

\item[Step 3.] Each time Bob receives a photon sent by Alice, he has no way of
knowing which quantum alphabet was chosen by Alice. \ So he simply uses the
flip of a fair coin to select one of the two alphabets and makes his
measurement accordingly. \ Half of the time he will be lucky and choose the
same quantum alphabet as Eve. In this case, the bit resulting from his
measurement will agree with the bit sent by Alice. \ However, the other half
of the time he will be unlucky and choose the alphabet not used by Alice. \ In
this case, the bit resulting from his measurement will agree with the bit sent
by Alice only 50\% of the time. After all these measurements, Bob now has in
hand a binary sequence $S_{Bob}$.
\end{itemize}

\bigskip

Alice and Bob now proceed to communicate over the public two-way channel using
the following stage 2 protocol:

\bigskip

\noindent\textbf{Stage 2 protocol: Communication over a public channel}%
\bigskip\ 

\begin{itemize}
\item[\textbf{Phase 1.}] \emph{Raw key extraction}\bigskip

\item[Step 1.] Over the public channel, Bob communicates to Alice which
quantum alphabet he used for each\ of his measurements.\bigskip

\item[Step 2.] In response, Alice communicates to Bob over the public channel
which of his measurements were made with the correct alphabet.\bigskip

\item[Step 3.] Alice and Bob then delete all bits for which they used
incompatible quantum alphabets to produce their resulting \textbf{raw keys}.
\ If Eve has not eavesdropped, then their resulting raw keys will be the same.
\ If Eve has eavesdropped, their resulting raw keys will not be in total agreement.

\item[\textbf{Phase 2.}] \emph{Error estimation}

\item[Step 1.] Over the public channel, Alice and Bob compare small portions
of their raw keys to estimate the error-rate $R$, and then delete the
disclosed bits from their raw keys to produce their \textbf{tentative final
keys}. \ If through their public disclosures, Alice and Bob find no errors
(i.e., $R=0$), then they know that Eve was not eavesdropping and that their
tentative keys must be the same \textbf{final key}. \ If they discover at
least one error during their public disclosures (i.e., $R>0$), then they know
that Eve has been eavesdropping. \ In this case, they discard their tentative
final keys and start all over again\footnote{If Eve were to intercept each
qubit received from Alice, to measure it, and then to masqurade as Alice by
sending on to Bob a qubit in the state she measured, then Eve would be
introducing a 25\% error rate in Bob's raw key. \ This method of eavesdropping
is called \textbf{opaque eavesdropping}. \ We will discuss this eavesdropping
strategy as well as others at a later time.}.
\end{itemize}%

\begin{align*}
&
\begin{tabular}
[c]{lllllllllll}\hline
\multicolumn{1}{|l}{Alice} & \multicolumn{1}{|l}{$\boxplus$} &
\multicolumn{1}{|l}{$\boxtimes$} & \multicolumn{1}{|l}{$\boxtimes$} &
\multicolumn{1}{|l}{$\boxtimes$} & \multicolumn{1}{|l}{$\boxplus$} &
\multicolumn{1}{|l}{$\boxtimes$} & \multicolumn{1}{|l}{$\boxplus$} &
\multicolumn{1}{|l}{$\boxtimes$} & \multicolumn{1}{|l}{$\boxplus$} &
\multicolumn{1}{|l|}{$\boxtimes$}\\\hline
\multicolumn{1}{|l}{} & \multicolumn{1}{|l}{$\updownarrow$} &
\multicolumn{1}{|l}{$\nwarrow$} & \multicolumn{1}{|l}{$\nwarrow$} &
\multicolumn{1}{|l}{$\nearrow$} & \multicolumn{1}{|l}{$\updownarrow$} &
\multicolumn{1}{|l}{$\nwarrow$} & \multicolumn{1}{|l}{$\leftrightarrow$} &
\multicolumn{1}{|l}{$\nearrow$} & \multicolumn{1}{|l}{$\leftrightarrow$} &
\multicolumn{1}{|l|}{$\nearrow$}\\\hline
\multicolumn{1}{|l}{} & \multicolumn{1}{|l}{1} & \multicolumn{1}{|l}{0} &
\multicolumn{1}{|l}{0} & \multicolumn{1}{|l}{1} & \multicolumn{1}{|l}{1} &
\multicolumn{1}{|l}{0} & \multicolumn{1}{|l}{0} & \multicolumn{1}{|l}{1} &
\multicolumn{1}{|l}{0} & \multicolumn{1}{|l|}{1}\\\hline
&  &  &  &  &  &  &  &  &  & \\\hline
\multicolumn{1}{|l}{Eve} & \multicolumn{1}{|l}{$\boxtimes$} &
\multicolumn{1}{|l}{$\boxplus$} & \multicolumn{1}{|l}{$\boxplus$} &
\multicolumn{1}{|l}{$\boxtimes$} & \multicolumn{1}{|l}{$\boxplus$} &
\multicolumn{1}{|l}{$\boxplus$} & \multicolumn{1}{|l}{$\boxtimes$} &
\multicolumn{1}{|l}{$\boxtimes$} & \multicolumn{1}{|l}{$\boxplus$} &
\multicolumn{1}{|l|}{$\boxplus$}\\\hline
\multicolumn{1}{|l}{} & \multicolumn{1}{|l}{1} & \multicolumn{1}{|l}{0} &
\multicolumn{1}{|l}{1} & \multicolumn{1}{|l}{1} & \multicolumn{1}{|l}{1} &
\multicolumn{1}{|l}{1} & \multicolumn{1}{|l}{0} & \multicolumn{1}{|l}{1} &
\multicolumn{1}{|l}{0} & \multicolumn{1}{|l|}{0}\\\hline
&  &  &  &  &  &  &  &  &  & \\\hline
\multicolumn{1}{|l}{Bob} & \multicolumn{1}{|l}{$\boxtimes$} &
\multicolumn{1}{|l}{$\boxtimes$} & \multicolumn{1}{|l}{$\boxplus$} &
\multicolumn{1}{|l}{$\boxtimes$} & \multicolumn{1}{|l}{$\boxplus$} &
\multicolumn{1}{|l}{$\boxtimes$} & \multicolumn{1}{|l}{$\boxplus$} &
\multicolumn{1}{|l}{$\boxplus$} & \multicolumn{1}{|l}{$\boxplus$} &
\multicolumn{1}{|l|}{$\boxplus$}\\\hline
\multicolumn{1}{|l}{} & \multicolumn{1}{|l}{1} & \multicolumn{1}{|l}{0} &
\multicolumn{1}{|l}{1} & \multicolumn{1}{|l}{1} & \multicolumn{1}{|l}{1} &
\multicolumn{1}{|l}{1} & \multicolumn{1}{|l}{1} & \multicolumn{1}{|l}{0} &
\multicolumn{1}{|l}{0} & \multicolumn{1}{|l|}{0}\\\hline
& $\ast$ & 0 & $\ast$ & 1 & 1 & 1 & 1 & $\ast$ & 0 & $\ast$\\
&  &  &  &  &  & E & E &  &  &
\end{tabular}
\\
&  \text{{\LARGE Fig 6b. \ The BB84 with\ Eve present (No noise)}}%
\end{align*}

\subsection{Quantum secrecy -- The BB84 protocol with noise}

Alice continues her presentation by addressing the issue of noise. \ 

\bigskip

``So far we have assumed that our cryptographic communication system is noise
free. \ But every realistic communication system has noise present.
\ Consequently, we now need to modify our quantum protocol to allow for the
presence of noise.''

\bigskip

``We must assume that Bob's raw key is noisy. \ Since Bob can not distinguish
between errors caused by noise and by those caused by Eve's intrusion, the
only practical working assumption he can adopt is that all errors are caused
by Eve's eavesdropping. \ Under this working assumption, Eve is always assumed
to have some information about bits transmitted from Alice to Bob. \ Thus, raw
key is always only \textbf{partially secret}.'' \ \bigskip

``What is needed is a method to distill a smaller secret key from a larger
partially secret key. \ We call this \textbf{privacy amplification}. \ We will
now create from the old protocol a new protocol that allows for the presence
of noise, a protocol that includes privacy amplification.''\bigskip

\noindent\textbf{Stage 1 protocol: Communication over a quantum channel}\bigskip

This stage is exactly the same as before, except that errors are now also
induced by noise.\bigskip

\noindent\textbf{Stage 2 protocol: Communication over a public channel}

\bigskip

\noindent\textbf{Phase 1 protocol: Raw key extraction.}

\begin{itemize}
\item[ ] 
\end{itemize}

This phase is exactly the same as in the noise-free protocol, except that
Alice and Bob also delete those bit locations at which Bob should have
received but did not receive a bit. \ Such ``non-receptions'' could be caused
by Eve's intrusion or by \textbf{dark counts} in Bob's detection device. \ The
location of dark counts are communicated by Bob to Alice over the public channel.

\bigskip

\noindent\textbf{Phase 2 protocol: Error estimation.}

\bigskip

Over the public channel, Alice and Bob compare small portions of their raw
keys to estimate the error-rate $R$, and then delete the disclosed bits from
their raw key to produce their \textbf{tentative final keys}. \ If $R$ exceeds
a certain threshold $R_{Max}$, then privacy amplification is not possible \ If
so, Alice and Bob return to stage 1 to start over. \ On the other hand, if
$R\leq R_{Max}$, then Alice and Bob proceed to phase 3.\bigskip

\noindent\textbf{Phase 3 protocol: Extraction of reconciled key\footnote{There
are more efficient and elegant procedures than the procedure descibed in Stage
2 Phase 3. See \cite{Lomonaco1} for references.}.}\bigskip

In this phase\footnote{The procedure given in Stage 2 Phase 3 is only one of
many different possible procedures. \ In fact, there are much more efficient
and elegant procedures than the one described herein.}, Alice and Bob remove
all errors from what remains of raw key to produce a common error-free key,
called \textbf{reconciled key}. \bigskip\ 

\begin{itemize}
\item[Step 1.] Alice and Bob publically agree upon a random permutation, and
apply it to what remains of their respective raw keys. \ Next Alice and Bob
partition the remnant raw key into blocks of length $\ell$, where the length
$\ell$ is chosen so that blocks of that length are unlikely to have more than
one error. \ For each of these blocks, Alice and Bob publically compare
overall parity checks, making sure each time to discard the last bit of each
compared block. \ Each time an overall parity check does not agree, Alice and
Bob initiate a binary search for the error, i.e., bisecting the block into two
subblocks, publically comparing the parities for each of these subblocks,
discarding the right most bit of each subblock. \ They continue their
bisective search on the subblock for which their parities are not in
agreement. \ This bisective search continues until the erroneous bit is
located and deleted. \ They then continue to the next $\ell$-block. \ 

This step is repeated, i.e., a random permutation is chosen, a remnant raw key
is partitioned into blocks of length $\ell$, parities are compared, etc..
\ This is done until it becomes inefficient to continue in this fashion.

\item[Step 2.] Alice and Bob publically select randomly chosen subsets of
remnant raw key, publically compare parities, each time discarding an agreed
upon bit from their chosen key sample. \ If a parity should not agree, they
employ the binary search strategy of Step 1 to locate and delete the error.
\end{itemize}

\bigskip

\begin{itemize}
\item Finally, when, for some fixed number $N$ of consecutive repetitions of
Step 2, no error is found, Alice and Bob assume that to a high probability,
the remnant raw key is without error. \ Alice and Bob now rename the remnant
raw key \textbf{reconciled key}, and proceed to the next phase.
\end{itemize}

\bigskip

\noindent\textbf{Phase 4: Privacy amplification}

Alice and Bob now have a common reconciled key which they know is only
partially secret from Eve. \ They now begin the process of \textbf{privacy
amplification}, which is the extraction of a secret key from a partially
secret one.\bigskip

\begin{itemize}
\item[Step 1.] Alice and Bob compute from the error-rate $R$ obtained in Phase
2 of Stage 2 an upper bound $k$ of the number of bits of reconciled key known
by Eve. \ 
\end{itemize}

\bigskip

Let $n$ denote the number of bits in reconciled key, and let $s$ be a
\textbf{security parameter} to be adjusted as required.

\bigskip

\begin{itemize}
\item[Step 2.] Alice and Bob publically select $n-k-s$ random subsets of
reconciled key, without revealing their contents. \ The undisclosed parities
of these subsets become the final secret key.
\end{itemize}

It can be shown that Eve's average information about the final secret key is
less than $2^{-s}/\ln2$ bits.

\bigskip

The bell rang, indicating the end of the period. \ The entire class with two
exceptions, immediately raced out of the lecture hall, almost knocking Alice
down as they passed by. \ Professor Dirac thanked Alice for an excellent
presentation. \ 

As Alice left, she saw Eve in one of the dark recesses of the large lecture
hall with her head resting on the palm of her hand as if in deep thought.
\ She had a frown on her face. \ Alice left with a broad smile on her face.

\section{The B92 quantum cryptographic protocol}

In the next class, Alice continued her last presentation. \ 

\bigskip

In thinking about the BB84 protocol this weekend, I was surprised to find that
it actually is possible to build a different quantum protocol that uses only
one quantum alphabet instead of two. \ I'll call this new quantum protocol
\textbf{B92}.''

``As before, we will describe the protocol in terms of the polarization states
of the photon\footnote{Any two dimensional quantum system such as a spin
$\frac{1}{2}$ particle could be used.}.''

\bigskip

``As our quantum alphabet, we choose
\[
\left\{
\begin{array}
[c]{ccc}%
``1" & = & \left|  \theta_{+}\right\rangle \\
&  & \\
``0" & = & \left|  \theta_{-}\right\rangle
\end{array}
\right.  \text{\qquad,}%
\]
where $\left|  \theta_{+}\right\rangle $ and $\left|  \theta_{-}\right\rangle
$ denote respectively the polarization states of a photon linearly polarized
at angles $\theta$ and $-\theta$ with respect to the vertical, where
$0<\theta<\frac{\pi}{4}$.''\bigskip

``We assume that Bob's quantum receiver, called a \textbf{POVM receiver}%
\cite{Brandt2}, is base on the following observables\footnote{The observables
$A_{\theta_{+}}$, $A_{\theta_{-}}$, and $A_{?}$ form a postive operator value
measure (POVM). \ }:
\[
\left\{
\begin{array}
[c]{lcl}%
A_{\theta_{+}} & = & \frac{1\ -\ \left|  \theta_{-}\right\rangle \left\langle
\theta_{-}\right|  }{1\ +\ \left\langle \theta_{+}\mid\theta_{-}\right\rangle
}\\
&  & \\
A_{\theta_{-}} & = & \frac{1\ -\ \left|  \theta_{+}\right\rangle \left\langle
\theta_{+}\right|  }{1\ +\ \left\langle \theta_{+}\mid\theta_{-}\right\rangle
}\\
&  & \\
A_{?} & = & 1-A_{\theta_{+}}-A_{\theta_{-}}%
\end{array}
\right.  \text{\quad,}%
\]
where $A_{\theta_{+}}$ is the observable for $\left|  \theta_{+}\right\rangle
$, $A_{\theta_{-}}$ the observable for $\left|  \theta_{-}\right\rangle $ and
$A_{?}$ is the observables for inconclusive receptions.''

\bigskip

The \textbf{B92} quantum protocol is as follows:

\bigskip

Stage 1 protocol. Communication over a quantum channel.

\bigskip

Step 1. The same as in the BB84 protocol. \ Alice flips a fair coin to
generate a random sequence $S_{Alice}$ of zeroes and ones. \ This sequence
will be used to construct a secret key shared only by Alice and Bob.\bigskip

Step 2. The same as in the previous protocol, except this time Alice uses only
one alphabet, the one above. \ So she does not have to flip a coin to choose
an alphabet.

\bigskip

Step 3. \ Bob uses his POVM receiver to measure photons received from Alice.

\bigskip

Stage 2. Communication in four phases over a public channel.

\bigskip

This stage is the same as in the BB84 protocol, except that in phase 1, Bob
publically informs Alice as to which time slots he received non-erasures.
\ The bits in these time slots become Alice's and Bob's raw keys.

\bigskip

Alice completed her discussion of the B92 protocol with,

\bigskip

``Eve's presence is again detected by an unusual error rate in Bob's raw key.
\ Moreover, for some but not all eavesdropping strategies, Eve can also be
detected by an unusual erasure rate for Bob.''\bigskip

Alice then stepped down from the lecture hall podium and returned to her seat.

\section{There are many other quantum cryptographic protocols}

Before continuing our story about Alice, Bob, and Eve, there are a few points
that need to be made:\bigskip

There are many other quantum cryptographic protocols. \ Quantum protocols
showing the greatest promise for security are those based on EPR pairs.
\ Unfortunately, the technology for implementing such protocols is not yet
available. \ For references on various protocols, please refer to
\cite{Lomonaco1}.

\section{A comparison of quantum cryptography with classical and public key cryptography}

\bigskip

Quantum cryptography's unique contribution is that it provides a mechanism for
eavesdropping detection. This is an entirely new contribution to cryptography.
On the other hand, one of the main drawbacks of quantum cryptography is that
it provides no mechanism for authentication, i.e., for detecting whether or
not Alice and Bob are actually communicating with each other, and not with an
intermediate Eve masquerading as each of them. \ Thus, the Catch 22 problem is
not solved by quantum cryptography. \ Before Alice and Bob can begin their
quantum protocol, they first need to send an authentication key over a secure channel.\bigskip

Thus, quantum cryptography's unique contribution is to provide a means of
expanding existing secure key. \ Quantum protocols are secure key expanders.
\ \ First a small authentication key is exchanged over a secure channel.
\ Then that key can be amplified to an arbitrary length through quantum cryptography.%

\[
\fbox{$%
\begin{array}
[c]{ll}%
\text{\underline{\textbf{Check List}} for Q. Crypto.\ Sys.} & \\
& \\
\qquad\blacksquare\text{ Catch 22 Solved?} & \qquad\qquad
\text{\textbf{YES\ \&\ NO}}\\
& \\
\qquad\blacksquare\text{ Authentication?} & \qquad\qquad\text{\textbf{NO}}\\
& \\
\qquad\blacksquare\text{ Intrusion Detection?} & \qquad\qquad
\text{\textbf{YES}}%
\end{array}
$}%
\]

\section{Eavesdropping strategies and counter measures}

\bigskip

Now let us resume our story:\bigskip

Not a split second after Alice had seated herself, Eve raised her hand and
asked for permission to make her own presentation to the class. \ Professor
Dirac yielded the podium, not knowing exactly what to expect, but nonetheless
elated that his usually phlegmatic class was beginning to show signs of
something he had not seen for some time, class participation and initiative.

Eve began, ``In the last two classes, Alice has suggested that I (Eve) might
be eager to eavesdrop on her conversations with my $\heartsuit$%
close$\heartsuit$ friend $\heartsuit$Bob$\heartsuit$. \ I assure you that that
simply is in no way true.'' \ 

\bigskip

``But such innuendo really doesn't bother me.''

\subsection{Opaque eavesdropping}

``What really irks me is that Alice suggests that, if I were to eavesdrop
(which never would happen), then I (Eve) would use \textbf{opaque
eavesdropping}. \ By \textbf{opaque eavesdropping}, I mean that I (Eve) would
intercept and observe (measure) Alice's photons, and then masquerade as Alice
by sending photons in the states I had measured on to Bob.''

``I assure you that, if I ever wanted to eavesdrop (which will never be the
case), I would not use such a simplistic form of intrusion.''

\bigskip

Eve really wanted to use the adjective `stupid' instead of `simplistic,' but
restrained herself.

\bigskip

Eve then said indignantly, ``If I ever were to eavesdrop (which would never
happen), I would use more sophisticated, more intelligent, and yes ... , more
deliciously devious schemes!''

\subsection{Translucent eavesdropping without entanglement}

``I (Eve) could for example make my probe interact unitarily with the
information carrier from Alice, and then let it proceed on to Bob in a
slightly modified state. For the B92 protocol, the interaction is given by:%

\[
\left\{
\begin{array}
[c]{c}%
\left|  \theta_{+}\right\rangle \left|  \psi\right\rangle \longmapsto U\left|
\theta_{+}\right\rangle \left|  \psi\right\rangle =\left|  \theta_{+}^{\prime
}\right\rangle \left|  \psi_{+}\right\rangle \\
\\
\left|  \theta_{-}\right\rangle \left|  \psi\right\rangle \longmapsto U\left|
\theta_{-}\right\rangle \left|  \psi\right\rangle =\left|  \theta_{-}^{\prime
}\right\rangle \left|  \psi_{-}\right\rangle
\end{array}
\right.  ,
\]
where $\left|  \psi\right\rangle $ and $\left|  \psi_{\pm}\right\rangle $
denote respectively the state of my (Eve's) probe before \ and after the
interaction and where $\left|  \theta_{\pm}\right\rangle $ and $\left|
\theta_{\pm}^{\prime}\right\rangle $ denote respectively the state of Alice's
photon before and after the interaction.''

\subsection{Translucent eavesdropping with entanglement}

``Another approach, one of the most sophisticated, would be for me (EVE) to
entangle my probe with the information carrier from Alice, \ and then let it
proceed on to Bob. For the B92 protocol, the interaction is given by:%

\[
\left\{
\begin{array}
[c]{c}%
\left|  \theta_{+}\right\rangle \left|  \psi\right\rangle \longmapsto U\left|
\theta_{+}\right\rangle \left|  \psi\right\rangle =a\left|  \theta_{+}%
^{\prime}\right\rangle \left|  \psi_{+}\right\rangle +b\left|  \theta
_{-}^{\prime}\right\rangle \left|  \psi_{+}\right\rangle \\
\\
\left|  \theta_{-}\right\rangle \left|  \psi\right\rangle \longmapsto U\left|
\theta_{-}\right\rangle \left|  \psi\right\rangle =b\left|  \theta_{+}%
^{\prime}\right\rangle \left|  \psi_{-}\right\rangle +a\left|  \theta
_{-}^{\prime}\right\rangle \left|  \psi_{-}\right\rangle
\end{array}
\right.  ,
\]
where $\left|  \psi\right\rangle $ and $\left|  \psi_{\pm}\right\rangle $
denote respectively the state of my (Eve's) probe before \ and after the
entanglement and where $\left|  \theta_{\pm}\right\rangle $ and $\left|
\theta_{\pm}^{\prime}\right\rangle $ denote respectively the state of Alice's
photon before and after the entanglement.''

\subsection{Eavesdropping based on implementation weaknesses}

``On the other hand, I could also take advantage of implementation weaknesses.''\bigskip

``One of the great difficulties with quantum cryptography is that technology
has not quite caught up with it. \ Many devices, such as lasers, do not emit a
single quantum, but many quanta at each emission time. \ The implementation of
quantum protocols really requires single-quantum emitters. \ Such
single-quantum emitters are now under development. \ Until such emitters
become available, the quantum protocols can only be approximately
implemented.'' \ \bigskip

``For example, for many optical implementations of quantum protocols, the
laser intensity is turned down so that on the average only one photon is
produced every 10 pulses. \ Thus, if anything is emitted at all (one chance
out of 10), then the probability that it is a single photon is extremely high.
\ However, when there is an emission, then there is a probability of
$\frac{1}{200}$ that more than one photon is emitted. \ So it is conceivable
that I (Eve) could build an eavesdropping device that would detect multiple
photon transmissions, and, when so detected, would divert one of the photons
for measurement. \ In this way, I (Eve) could conceivably read $\frac{1}{200}$
of Alice's transmission without being detected. \ \ One way of countering this
type of threat is to allow for it during privacy amplification. \ Another is
to develop devices which actually truly emit one quanta at a time.''\bigskip

``Finally, depending on Alice's implementation, it might also be possible for
me (Eve) to gain information simply by observing Alice's transmitter without
measuring its output. \ This may or may not be far fetched.''

\bigskip

Eve then returned to her seat. \ Her face was lit up with a sinister grin of satisfaction.

\section{Implementations}

Before continuing our story, we should mention that quantum cryptographic
protocols have been implemented over more than 30 kilometers of fiber optic
cable, \cite{Phoenix2},\cite{Townsend1},\cite{Townsend2}, \cite{Townsend3},
and most amazingly, over more than a kilometer of free space \cite{Buttler1},
\cite{Buttler2}, \cite{Jacobs1},\cite{Franson2}, \cite{Hughes1} in the
presence of ambient sunlight. \ There have been a number of ambitious
proposals to demonstrate the feasibility of quantum cryptography in earth to
satellite communications. \ And as mentioned earlier, there is a clear need
for the development of single-quantum emitting devices. \ 

\section{Conclusion}

\bigskip

Much remains to be done. \ There has been some work on the development of
multiple-user quantum cryptographic protocols for communication
networks\cite{Townsend6}. \ There also have been at least two independent
claims of the proof of ultimate security, i.e., a proof that quantum
cryptographic protocols are impervious to all possible eavesdropping
strategies \cite{Lo3}, \cite{Mayers2}, \cite{Mayers3}, \cite{Mayers4}.\bigskip

Our story continues:\bigskip

As Alice sat in her seat, she happened to spy in the corner of her eye an
abrupt change in Eve's demeanor. \ Eve suddenly became agitated, lit up with
excitement, and started to frantically write on her notepad. \ The bell rang.
\ Eve immediately jumped up, and raced out of the lecture hall, being pushed
along by the usual frantic mass of students, equally eager to get out of the
classroom. \ 

As Eve whisked past, Alice caught just a fleeting glimpse of Eve's notepad.
\ All Alice was able to discern in that brief moment was an illegible jumble
of equations and ... yes, ... the acronym ``POVM.''

\bigskip

Alice thought to herself, ``Oh, well! ... Forget it! \ I think I'll just visit
Bob this weekend.''

\bigskip

\begin{center}
{\huge THE END}\footnote{Any resemblance of the characters in this manuscript
to individuals living or dead is purely coincidental.}
\end{center}

\bigskip

\section{Acknowledgement}

I would like to thank Howard Brandt and Lov Grover for their helpful
suggestions. \ I would also like to thank\ the individuals who attended my
talk. \ Their many comments and insights were of invaluable help in writing
this paper. \ Thanks are also due to the NIST\ Computer Security Division for
providing an encouraging environment in which this paper could be completed.

\end{document}